\documentclass{jfm}

\pdfoutput=1

\usepackage{graphicx}
\usepackage{xcolor}
\usepackage{natbib}
\usepackage{amsmath}
\usepackage{amssymb}
\usepackage{bm}

\newcommand{\etal}{et al.\ }
\newcommand{\ie}{i.e.,\ }
\newcommand{\eg}{e.g.,\ }

\newcommand{\figrefC}[2][]{\mbox{Figure \ref{#2}(#1)}}
\newcommand{\figrefAC}[1]{\mbox{Figure \ref{#1}}}
\newcommand{\tabrefC}[1]{\mbox{Table \ref{#1}}}

\newcommand{\equref}[1]{\mbox{equation (\ref{#1})}}
\newcommand{\figref}[2][]{\mbox{figure \ref{#2}(#1)}}
\newcommand{\figrefA}[1]{\mbox{figure \ref{#1}}}

\newcommand{\secref}[1]{\mbox{section \ref{#1}}}

\title[Turbulent duct flow with polymers]
{Turbulent duct flow with polymers}

\author[A. Shahmardi \etal]{Armin Shahmardi$^1$, Sagar Zade$^1$, Mehdi N. Ardekani$^1$, Rob J. Poole$^2$, Fredrik Lundell$^1$, Marco E. Rosti$^1$\thanks{Email address for correspondence: merosti@mech.kth.se}, and  Luca Brandt$^1$}

\affiliation{$^1$ Linn\'{e} Flow Centre and SeRC (Swedish e-Science Research Centre), \\KTH Mechanics, SE 100 44 Stockholm, Sweden \\ $^2$ Department of Engineering, University of Liverpool, \\ Brownlow Street, Liverpool, L69 3GH United Kingdom}

\begin{document}

\maketitle

\begin{abstract}
We have performed direct numerical simulation of the turbulent flow of a polymer solution in a square duct, with the FENE-P model used to simulate the presence of polymers. First, a simulation at a fixed moderate Reynolds number is performed and its results compared with those of a Newtonian fluid to understand the mechanism of drag reduction and how the secondary motion, typical of the turbulent flow in non-axisymmetric ducts, is affected by polymer additives. Our study shows that the Prandtl's secondary flow is modified by the polymers: the  circulation of the streamwise main vortices increases and the location of the maximum vorticity move towards the center of the duct. In-plane fluctuations are reduced while the streamwise ones are enhanced in the center of the duct and dumped in the corners due to a substantial modification of the quasi-streamwise vortices and the associated near-wall low- and high-speed streaks; these grow in size and depart from the walls, their streamwise coherency increasing. Finally, we investigated the effect of the parameters defining the viscoelastic behavior of the flow and found that the Weissenberg number strongly influences the flow, with the cross-stream vortical structures growing in size and the in-plane velocity fluctuations reducing for increasing flow elasticity.
\end{abstract}

\section{Introduction} \label{sec:introduction}
\subsection{Aim and objectives}
Near-wall turbulence is responsible for significant drag penalties in many flows of engineering relevance, and because of that, many researchers are studying various ways to be able to properly control the flow \citep{choi_moin_kim_1993a, dubief_white_terrapon_shaqfeh_moin_lele_2004a, breugem_boersma_uittenbogaard_2006a, orlandi_leonardi_2008a, garcia-mayoral_jimenez_2011b, rosti_cortelezzi_quadrio_2015a, rosti_brandt_pinelli_2018a}. Among the many control strategies, the use of polymers has been demonstrated to be very efficient to reduce drag in pipelines \citep{virk_1971a}. While flows through axisymmetric geometry have been extensively studied \citep{virk_merrill_mickley_smith_mollo-christensen_1967a, cho_harnett_1982a, den-toonder_hulsen_kuiken_nieuwstadt_1997a, escudier_presti_smith_1999a, ptasinski_boersma_nieuwstadt_hulsen_van-den-brule_hunt_2003a, resende_escudier_presti_pinho_cruz_2006a}, less attention has been given to more complex geometries, such as square ducts. In this context, the aim of this work is to explore and better understand the interactions between the turbulent flow in ducts with square section and polymer additives.

\subsection{Duct flow}
Turbulent flow in a plane channel has been extensively studied in the past \citep{kim_moin_moser_1987a} while the flow in a duct with a square cross-section has received much less attention \citep{uhlmann_pinelli_kawahara_sekimoto_2007a, pinelli_uhlmann_sekimoto_kawahara_2010a, samanta_vinuesa_lashgari_schlatter_brandt_2015a, vinuesa_schlatter_nagib_2015a, owolabi_poole_dennis_2016a}, although its geometry is only mildly more complex. Peculiar features of the turbulent flow in a duct are that the mean velocity profile presents a non-uniform distribution of the skin friction coefficient along the edges, and the related appearance in the cross-sectional plane of secondary motions of the second kind, as classified by \citet{prandtl_1926a}, which is a mean flow effect induced by gradients of turbulence fluctuations, \eg a breaking of axisymmetry. The existence of such secondary mean motion in this geometrical configuration is well known since the experiments by \citet{nikuradse_1926a}, which was followed by further experimental measurements \citep{brundrett_baines_1964a, gessner_1973a, melling_whitelaw_1976a} as well as direct and large eddy simulations \citep{madabhushi_vanka_1991a, gavrilakis_1992a, uhlmann_pinelli_kawahara_sekimoto_2007a, pinelli_uhlmann_sekimoto_kawahara_2010a} all of which subsequently extended our knowledge of this flow. \citet{uhlmann_pinelli_kawahara_sekimoto_2007a} simulated the case of the marginal Reynolds number regime  and showed that the buffer layer coherent structures play a crucial role in the appearance of the secondary flow and in the deformation of the mean streamwise velocity profile. Indeed, they proposed that the deformation of the mean streamwise velocity profile is due to the presence of preferential positions of quasi-streamwise vortices and velocity streaks. Moreover, in such a marginal regime, short-time averaged velocity fields are found to exhibit a 4-vortex state instead of the usual 8-vortex secondary flow pattern found at higher Reynolds numbers. This feature was explained with the relation between coherent structures and secondary flow: if the dimension of the cross-section, in wall units, is below the one needed to accommodate a complete minimal turbulent cycle on all four walls \citep{jimenez_moin_1991a}, then just two facing walls can alternatively give rise to a complete turbulent regeneration mechanism, while the other two faces remain in a relative quiescent state. Such 4-vortex states at marginal Reynolds number have also been observed experimentally \citep{owolabi_poole_dennis_2016a}. \citet{pinelli_uhlmann_sekimoto_kawahara_2010a} studied turbulent duct flows at higher Reynolds numbers, where the length of the edge of the square cross-section, expressed in viscous units, is larger, therefore allowing for the simultaneous presence of multiple pairs of high- and low-velocity streaks. These authors further proved the idea that the secondary flow is a footprint of the coherent motions embedded in the turbulent flow, and proposed a scenario in which the position of the center of the mean secondary vortices is determined by the preferential positioning of the quasi-streamwise vortices.

\subsection{Polymer additives}
Polymer addition is a very efficient strategy employed for drag reduction in wall-bounded turbulent flows, since drag reduction (DR) up to $80\%$ have been achieved with only few ppm concentrations. Two distinct regimes are usually identified \citep{warholic_massah_hanratty_1999a}, often called low and high drag reduction, LDR and HDR, respectively. The former (LDR) exhibits a log-law region of the mean velocity profile parallel to that of the Newtonian flow but with an upward shift associated with the amount of drag reduction; streamwise fluctuations are increased while wall-normal and spanwise components reduced, as well as the shear stress. The regime characterised by high drag reductions (HDR), more than $40\%$, shows a strong increase in the slope of the log-law region and low levels of Reynolds shear stress \citep{ptasinski_boersma_nieuwstadt_hulsen_van-den-brule_hunt_2003a}. The drag reduction is eventually bounded by a maximum (MDR) \citep{virk_mickley_smith_1970a}. Several works \citep{warholic_massah_hanratty_1999a, ptasinski_boersma_nieuwstadt_hulsen_van-den-brule_hunt_2003a} observed low Reynolds shear stress and a consequent deficit in the stress balance equation, which has been interpreted as the input of energy from the polymers to the flow which ultimately sustains the asymptotic MDR.

The drag reduction mechanism is quite complex due to its multiscale nature (small amount of microscopic polymer molecules is needed to achieve significant drag reduction in the bulk flow) and its full physical understanding is still incomplete. Several explanations have been proposed to gain further insight in the mechanism of polymer drag reduction: Dimitropoulos et al. (2001) showed that streamwise enstrophy is inhibited by the extensional viscosity generated by polymers stretched by turbulence; Ptasinski et al. (2003) found a shear sheltering effect in the near-wall region which decouples the outer and inner layer vortices; Min, Yoo \& Choi (2003a) and Min et al. (2003b) observed the transport of elastic energy from the viscous sublayer to the buffer and the log region, and proposed a criterion for the onset of drag reduction. 

\citet{dubief_white_terrapon_shaqfeh_moin_lele_2004a, dubief_terrapon_white_shaqfeh_moin_lele_2005a} studied the intermittency of polymers in turbulent flows, showing that the action of polymers is as intermittent as the near-wall vortices, and that the drag-reducing property of polymers is closely related to coherent turbulent structures. Polymers dampen near-wall vortices but also enhance streamwise kinetic energy in near-wall streaks. The net balance of these two opposite actions leads to a self-sustained drag-reduced turbulent flow: the polymers reduce turbulence by opposing the downwash and upwash flows generated by near-wall vortices, while they enhance streamwise velocity fluctuations in the very near-wall region. Recent studies by \citet{xi_graham_2010a, xi_graham_2012a, xi_graham_2012b} provided new insight on the mechanism by which polymer additives reduce the drag. These authors suggested that a turbulent flow is characterised by an alternate succession of strong and weak turbulence phases. The first are characterised by flow structures showing strong vortices and wavy streaks, the latter weak streamwise vortices and almost streamwise-invariant streaks. In the Newtonian flow, the so-called active turbulence dominates; with increasing viscoelasticity, on the contrary, active intervals becomes shorter while the so-called hibernating intervals are unaffected. Also, it is shown that during these hibernating turbulence intervals, the turbulent dynamics resemble MDR turbulence in both Newtonian and viscoelastic flows \citep{li_sureshkumar_khomami_2006a, white_somandepalli_mungal_2004a}.

Polymers can also alter flow instabilities and transition to turbulence. Indeed, \citet{biancofiore_brandt_zaki_2017a} recently examined the secondary instability of streaks in a viscoelastic flow, showing that the streaks reach a lower average energy with increasing elasticity due to a resistive polymer torque that opposes the streamwise vorticity and, as a result, opposes the lift-up mechanism.

\subsection{Polymer solutions in square duct flows}
Extensive work has been done experimentally to try to characterise polymer solution behaviour in duct flows. In particular, \citet{rudd_1972a} and \citet{logan_1972a} reported limited measurements of the mean-flow and turbulence structure for flow of drag-reducing polymers through square tubes, with sufficiently low polymer concentrations, and provided no information about the secondary flow. More recently, \citet{gampert_rensch_1996a} have reported the results of a systematic study of the influence of polymer concentration on near-wall turbulence structure for the flow through a square duct, but again have provided no direct information on the secondary flow. Also, they argued that there are two flow regimes in which the properties of the polymer structure, and hence the turbulent flow-field, differ significantly, depending on the polymer concentration. \citet{escudier_smith_2001a} presented both global data (friction-factor versus Reynolds number) and detailed mean axial- and secondary-flow-velocity and turbulence-­field data for fully-developed flow through a square duct of various polymer solutions. They found a reduction in the intensity of turbulent-velocity fluctuations transverse to the mean flow and a strong reduction in the secondary-flow velocities. More recently, \citet{escudier_nickson_poole_2009a} provided a comprehensive experimental database of previous experiental works \citep{rudd_1972a, gampert_yong_1990a, escudier_smith_2001a, gampert_braemer_eich_dietmann_2005a}; the authors pointed out that several of these studies have limited turbulence data, and considers mainly relatively low polymer concentrations. \citet{escudier_nickson_poole_2009a} gave special emphasis on the quantification of turbulence anisotropy and showed that with polymer additives, the flow displays a tendency towards the axisymmetric-turbulence limit. Also, there is a marked decrease in anisotropy with distance from the near-surface peak in all cases but this tendency progressively reduces with increasing concentration/drag-reduction level. Recently, \citet{owolabi_dennis_poole_2017a} have explored the relationship between drag reduction and fluid elasticity, and describe a technique for an a priori quantitative prediction of drag reduction from a knowledge of polymer relaxation time, flow rate and geometric length scale, based on a universal relationship between drag reduction and fluid elasticity. Although the previous discussion suggests a satisfactory situation from an engineering view-point, for polymer solutions a full understanding of the phenomena is still missing.

\subsection{Outline}
In this work, we present Direct Numerical Simulations (DNS) of a turbulent duct flow with polymer additives at moderate Reynolds number. In the simulations, the viscoelastic fluid is modelled by the constitutive FENE-P (Finite Extensible Nonlinear Elasticity-Peterlin) closure. In \secref{sec:formulation}, we first discuss the flow configuration and governing equations, and then present the numerical methodology used. The features of the turbulent duct flow are documented in \secref{sec:result}, where we compare results with and without the polymer additives, to elucidate their effect on the secondary motion and on the flow in general at this moderate Reynolds number. First and second order statistics are analysed, together with energy and vorticity budgets. Finally, a summary of the main findings and some conclusions are drawn in \secref{sec:conclusion}.

\section{Formulation} \label{sec:formulation}
\begin{figure}
  \centering
  \includegraphics[width=0.4\textwidth]{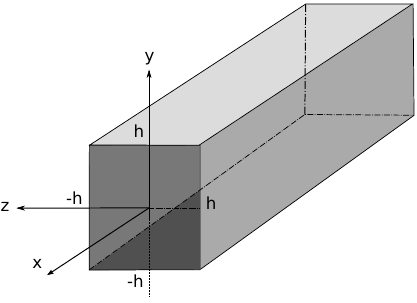}
  \caption{Sketch of the square duct geometry and coordinate system adopted with the walls located at $y=-h$ and $y=h$, and at $z=-h$ and $z=h$.}
  \label{fig:sketch}
\end{figure}
We consider the turbulent flow of an incompressible viscoelastic fluid through a square duct. \figrefAC{fig:sketch} shows a sketch of the geometry and the Cartesian coordinate system, where $x$, $y$ and $z$ ($x_1$, $x_2$, and $x_3$) denote the streamwise, and the two cross-stream coordinates, while $u$, $v$ and $w$ ($u_1$, $u_2$, and $u_3$) denote the respective components of the velocity vector field. The lower, upper walls are located at $y=\pm h$, while the left and right walls at $z=\pm h$, where $h$ is the cross section half size. Moreover, we define $\widetilde{y}$ as the distance from the wall. The Reynolds number of the flow is defined as $Re = U_b h/\nu$, where $h$ is chosen as characteristic length scale, the characteristic velocity is the bulk velocity $U_b$, defined as the average value of the mean velocity computed across the whole domain, and $\nu$ the total kinematic viscosity, \ie the sum of the solvent and polymer viscosity. Hereafter, all the velocity components will be made non-dimensional with $U_b$ and lengths with $h$, except where explicitly stated otherwise.

The fluid is governed by the incompressible Navier-Stokes (NS) equations:
\begin{subequations}
\label{eq:NS}
\begin{align}
\frac{\partial u_i}{\partial t} + \frac{\partial u_i u_j}{\partial x_j} &= - \frac{\partial p}{\partial x_i} + \frac{\beta}{Re} \frac{\partial^2 u_i}{\partial x_j \partial x_j} + \frac{1-\beta}{Re} \frac{\partial \tau_{ij}}{\partial x_j}, \\
\frac{\partial u_i}{\partial x_i} &= 0,
\end{align}
\end{subequations}
where $p$ is the pressure, $\beta$ the ratio of the solvent viscosity to the total fluid viscosity, and $\tau_{ij}$ the extra stress tensor due to the polymers. Note that, if $\beta$ equals $1$, the standard Navier-Stokes equations are recovered. To model the additional stresses due to the presence of polymers in the flow, we use the FENE-P closure where the polymer stress tensor $\tau_{ij}$ is written as a function of the configuration tensor $C_{ij}$ as
\begin{equation} \label{eq:polymerStress}
\tau_{ij}=\frac{1}{Wi}\left(\frac{C_{ij}}{1-\frac{C_{kk}}{L^2}}-\delta_{ij}\right);
\end{equation}
in the expression above, $L$ is the dumbbell extensibility, $C_{kk}$ the trace of the configuration tensor, $Wi$ the Weissenberg number defined as the ratio between the elastic and viscous forces \citep{dealy_2010a, poole_2012a}, \ie $Wi=\lambda U_b/h$ being $\lambda$ the polymer relaxation time, and $\delta_{ij}$ the Kronecker delta. The configuration tensor is a symmetric second order tensor, which is found by solving the following dynamic equation:
\begin{equation}\label{eq:polymerDynamic}
\frac{\partial {C_{ij}}}{\partial{t}} + u_k\frac{\partial {C_{ij}}}{\partial{x_k}}=C_{kj}\frac{\partial {u_{i}}}{\partial{x_k}}+C_{ik}\frac{\partial {u_{j}}}{\partial{x_k}}-\tau_{ij}.
\end{equation}
The equation is a balance between the advection of the configuration tensor on the left-hand side, and the stretching and relaxation of the polymer, represented by the first two terms and the last one on the right-hand side, respectively.

\subsection{Numerical implementation}
The code for the numerical solution of the equations above is an extension of that described by \citet{rosti_brandt_2017a,rosti_brandt_2018a,rosti_brandt_mitra_2018a}, with whom it shares its general architecture. The time integration method used to solve the equations is based on an explicit fractional-step method \citep{kim_moin_1985a}, where all the terms are advanced with the third order Runge-Kutta scheme, except the viscous stress contribution which is advanced with Crank-Nicolson \citep{min_yoo_choi_2001a}. In particular, to solve the system of governing equations, at every time step, we perform the following steps \citep[see also][]{dubief_terrapon_white_shaqfeh_moin_lele_2005a, min_yoo_choi_2001a}: i) the updated conformation tensor $C_{ij}$ is found by solving \equref{eq:polymerDynamic} with the last term $\tau_{ij}$ substituted by \equref{eq:polymerStress}; ii) the polymer stress tensor $\tau_{ij}$ is computed by \equref{eq:polymerStress}; iii) the NS equations (\equref{eq:NS}) are advanced in time by first solving the momentum equation (prediction step), then by solving a Poisson equation for the projection variable, and finally by correcting the velocity and pressure to make the velocity field divergence free (correction step). 

The numerical solution of \equref{eq:polymerDynamic} is cumbersome, and many reaserchers showed that the numerical solution of a viscoelastic fluid is unstable, especially in the case of high Weissenberg numbers, since any disturbance amplifies over time \citep{dupret_marchal_1986a, sureshkumar_beris_1995a,min_yoo_choi_2001a}. Indeed, \equref{eq:polymerDynamic} can easily diverge and lead to the numerical breakdown of the solution since it is an advection equation without any diffusion term \citep{dubief_terrapon_white_shaqfeh_moin_lele_2005a}. One of the earliest solution to this problem has been to introduce an artificial diffusivity (AD) to the constitutive equations \citep{sureshkumar_beris_1995a, mompean_deville_1997a, alves_pinho_oliveira_2000a}. Subsequently, global AD was replaced by local AD, a method which has been widely used. In the present work, we use global AD in the first transient part of the simulation of a viscoelastic flow, and then gradually remove it while approaching the final statistical steady state regime, when it is completely removed. Also, we chose to use a fifth order Weighted Essentially Non-Oscillatory (WENO) scheme \citep{jiang_shu_1996a} for the advection terms in \equref{eq:polymerDynamic}. Apart from that, the governing differential equations are solved on a staggered grid using a second order central finite-difference scheme. This methodolgy has been proved to work properly by \cite{sugiyama_ii_takeuchi_takagi_matsumoto_2011a} and also successfully used in \citet{rosti_brandt_2017a} for viscoelastic solid-like materials. A comprehensive review on the properties of different numerical schemes for the advection term is reported by \cite{min_yoo_choi_2001a}.

\subsection{Numerical details}
For all the turbulent flows considered hereafter, the equations of motion are discretised by using $448 \times 448 \times 448$ grid points on a computational domain of $12h \times 2h \times 2h$ in the streamwise, wall-normal and spanwise directions, respectively. The spatial resolution has been chosen in order to properly resolve the wall turbulence, satisfying the constraint $ \Delta y^+ = \Delta z^+ \approx 0.8$ and $\Delta x^+ \approx 5$ in the Newtonian case. 

Viscous units, used above to express spatial resolution, will be often employed in the following; they are indicated by the superscript $^+$, and are built using the friction velocity $u_\tau$ as velocity scale and the viscous length $\delta_\nu = \nu / u_\tau$ as length scale. For a turbulent Newtonian duct flow, the friction velocity is defined as
%u_\tau = \sqrt{\frac{1}{Re} \frac{\partial \overline{u}}{\partial y} },
\begin{equation} 
\label{eq:friction_velocity}
u_\tau = \sqrt{\nu \frac{\partial u^*}{\partial y^*} },
\end{equation}
where $\nu$ is the kinematic viscosity, $u^*$ and $y^*$ denote the dimensional values of velocity and distance from wall , and the derivative is taken at the wall (note that, the equation is written for the top and bottom wall, but a similar one can be written for the left and right by changing $y$ with $z$). When the flow is viscoelastic, the definition \eqref{eq:friction_velocity} must be modified to account for the polymeric stress which is in general non-zero at the wall. Thus, we define
\begin{equation} \label{eq:friction_velocity_total}
u_\tau = \sqrt{\beta \nu \frac{d u^*}{d {\tilde{y}}^*} + \left( \frac{1-\beta}{\rho} \right) {\tau_{12}}^*},
\end{equation}
being ${\tau_{12}}^*$ the dimensional polymeric stress.
All the simulations are performed at constant flow rate consistently with choosing $U_b$ as the characteristic velocity; the flow Reynolds number based on the bulk velocity is fixed at $2800$, \ie $Re=U_b h/\nu=2800$, where the bulk velocity is the average value of the mean velocity computed across the whole domain occupied by the fluid phase. This choice facilitates the comparison between the Newtonian and viscoelastic flow. Since we are enforcing the constant flow rate condition, the appropriate instantaneous value of the streamwise pressure gradient is determined at every time step.

All the simulations are started from a fully developed turbulent duct flow without polymer additives. After the flow has reached statistical steady state, the calculations are continued for an interval of $400h/U_b$ time units, during which $400$ full flow fields are stored for further statistical analysis. To verify the convergence of the statistics, we have computed them using different number of samples and verified that the differences are negligible.

\subsection{Code validation}
\begin{figure}
\centering
\includegraphics[width = 0.325\textwidth]{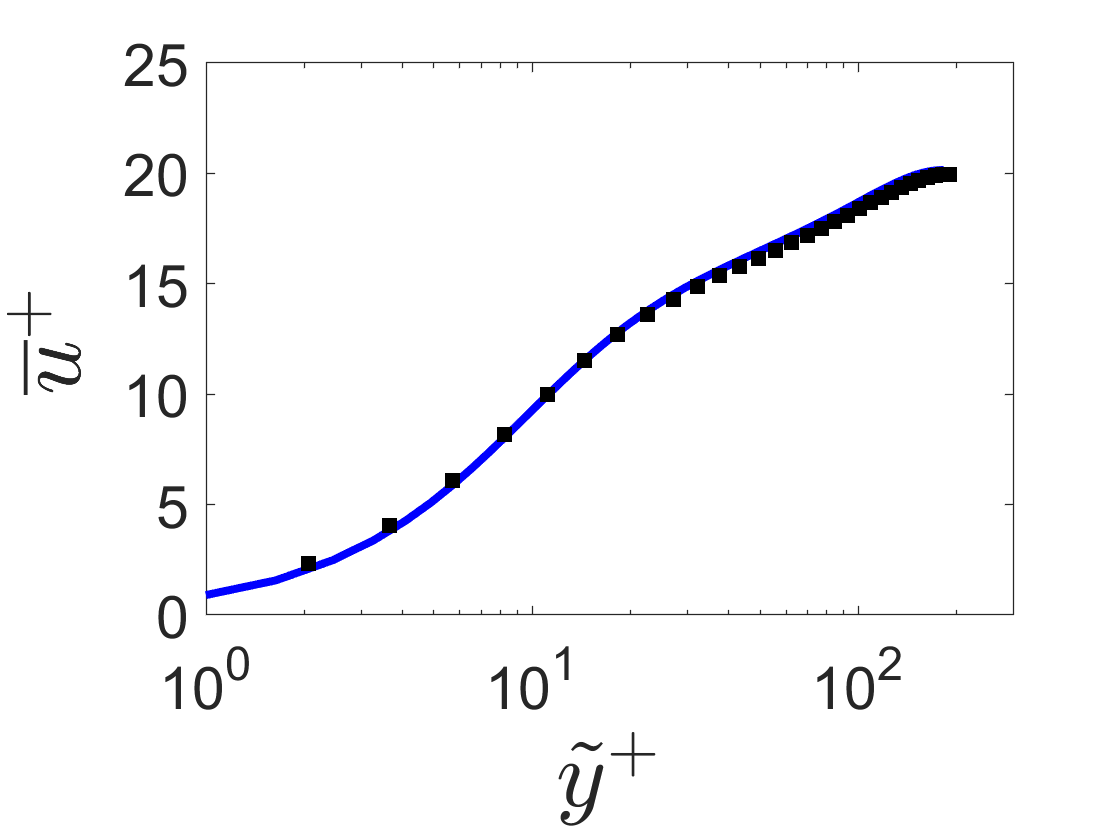}
\includegraphics[width = 0.325\textwidth]{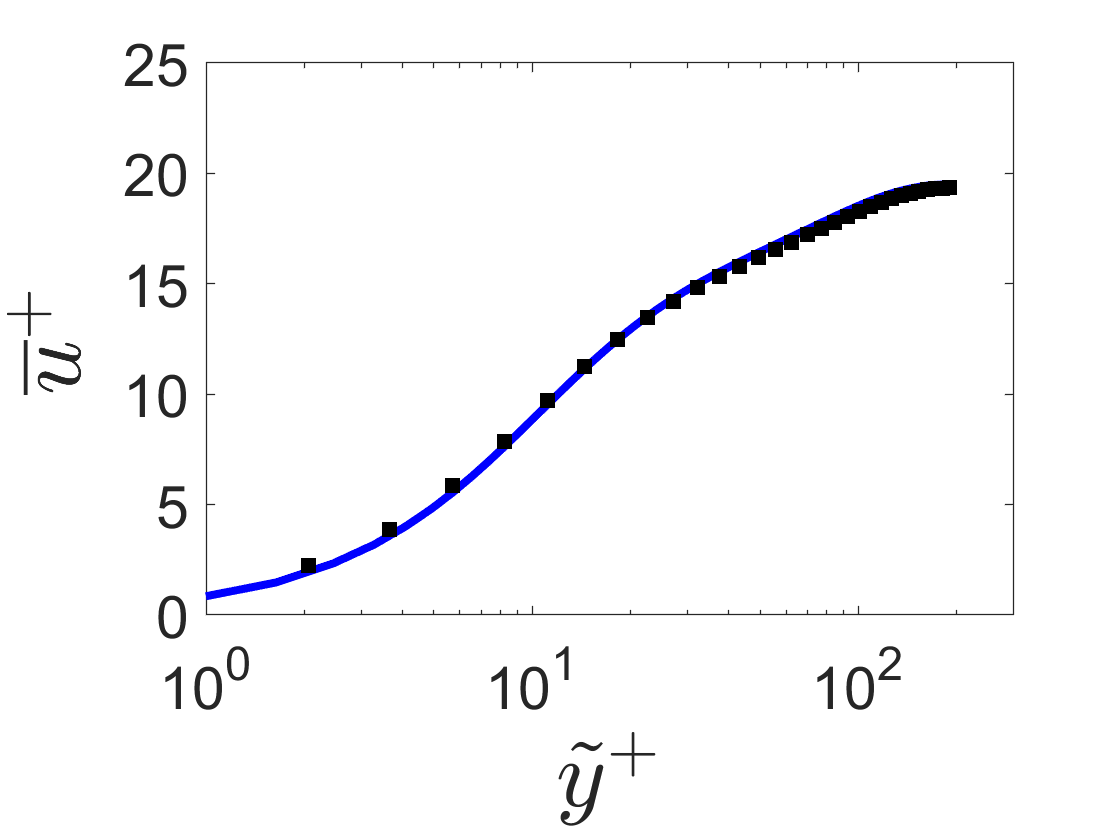}
\includegraphics[width = 0.325\textwidth]{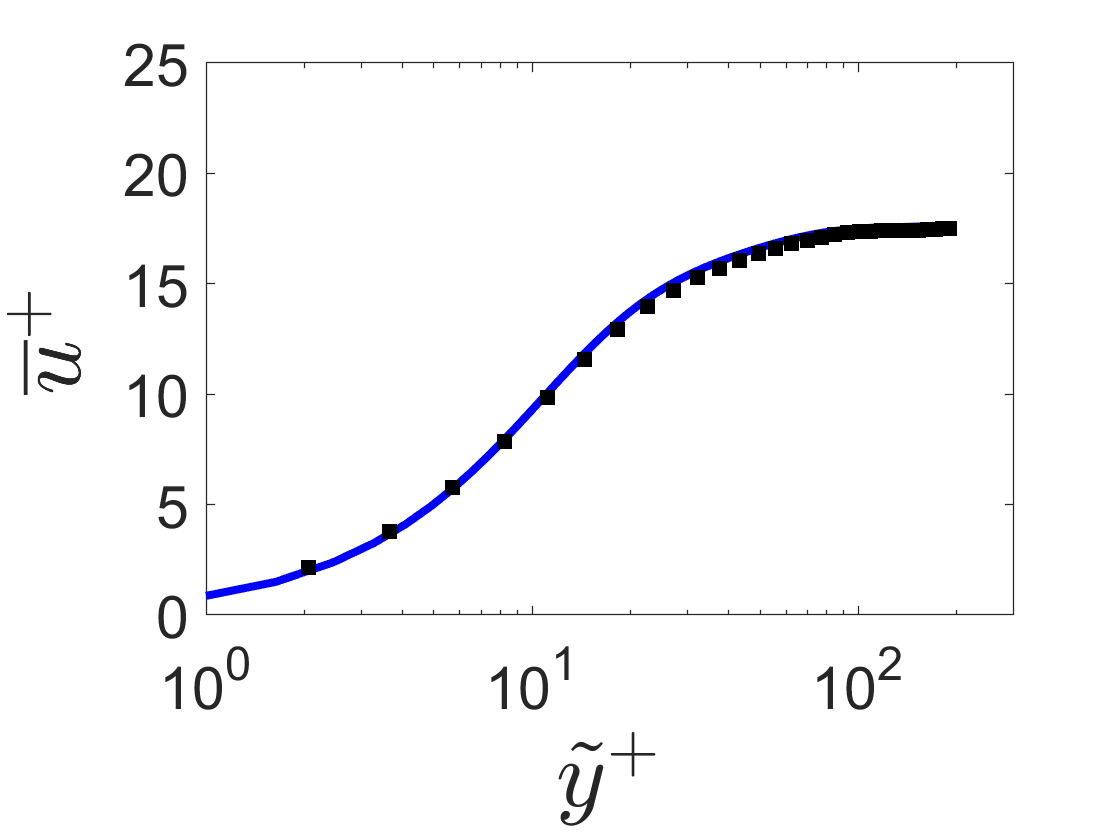}
\includegraphics[width = 0.325\textwidth]{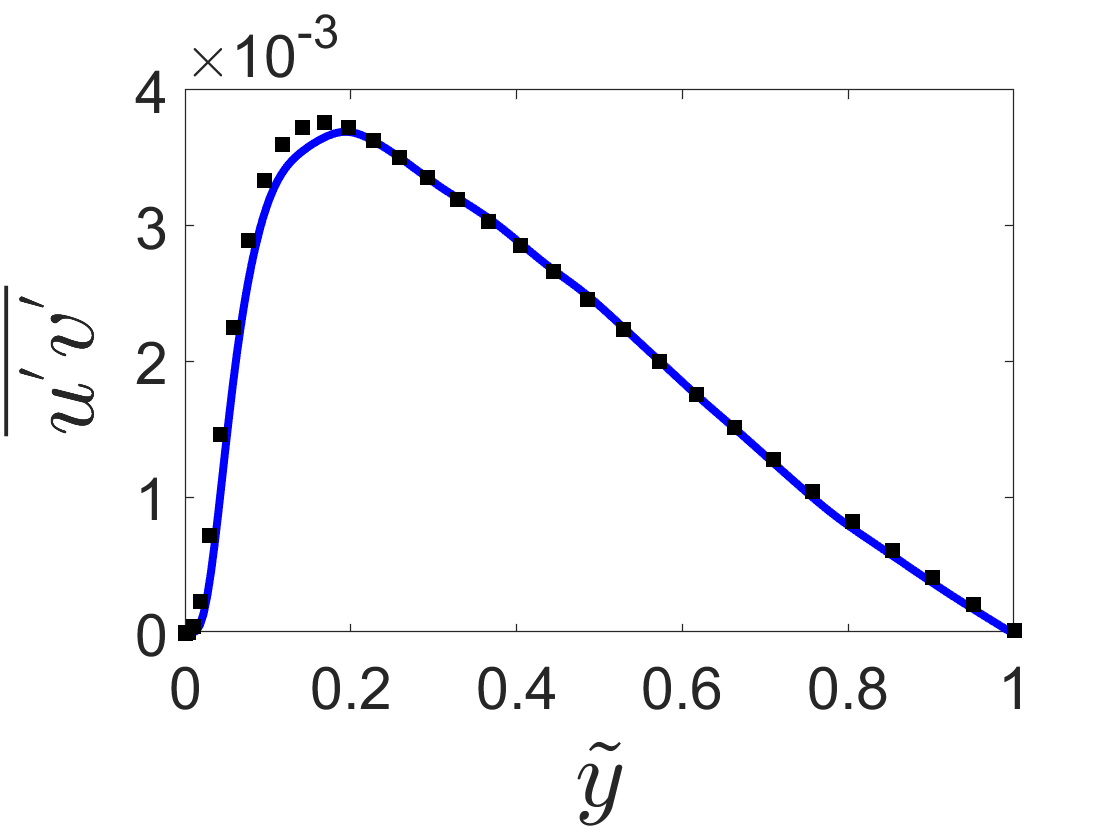}
\includegraphics[width = 0.325\textwidth]{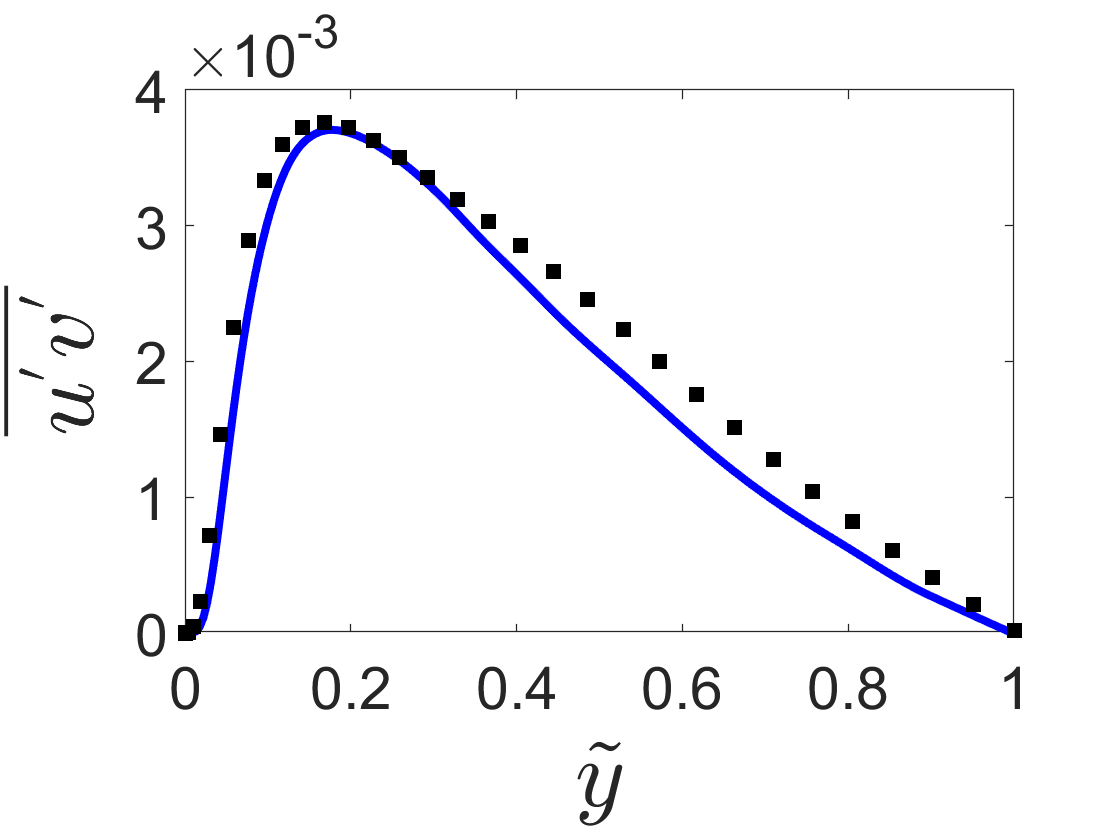}
\includegraphics[width = 0.325\textwidth]{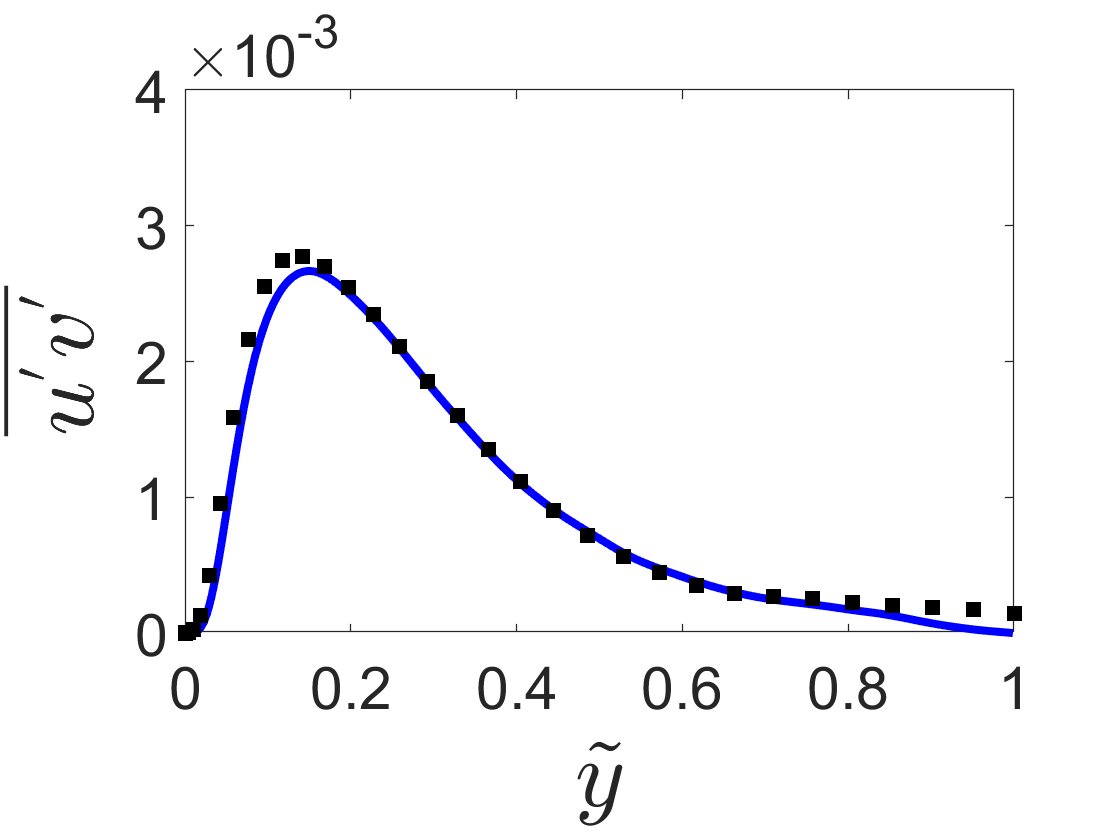}
\caption{(top) Mean streamwise velocity component $u$ as a function of the wall normal distance, indicated with $\tilde{y}$, for the case of a Newtonian turbulent duct flow. (bottom) Mean $u'v'$ component of the Reynolds stress tensor profile in bulk units. In the figures, the left, middle and right panels correspond to the section $z=0$, $0.3h$ and $0.6h$, respectively. Solid lines are used for our numerical results, while the symbols are the results those from \citet{uhlmann_pinelli_kawahara_sekimoto_2007a} and \citet{pinelli_uhlmann_sekimoto_kawahara_2010a}.}
\label{fig:val-num}
\end{figure}

The code has been extensively validated in the past for turbulent flow simulations \citep{rosti_brandt_2017a}. Here, we provide one more comparison with literature results for the case of a turbulent duct flow. We compare our results of a Newtonian turbulent duct flow at $Re=2800$ with those reported by \citet{uhlmann_pinelli_kawahara_sekimoto_2007a} and \citet{pinelli_uhlmann_sekimoto_kawahara_2010a} at a slightly higher Reynolds number $Re=2900$. \figrefAC{fig:val-num} shows the mean velocity profile in logarithmic scale in the top row and the cross term of the Reynolds stress tensor $u'v'$ as a function of the distance from the wall,  $\tilde{y}$. Three different section are considered: $z=0$ the centerline, $z=0.3h$ and $z=0.6h$. Overall good agreement between the literature (shown with the black symbols) and our results (shown with the solid line) is evident.

The FENE-P model implementation has been validated by simulating a viscoelastic temporally evolving mixing layer flow and by comparing our results with those provided by \citet{min_yoo_choi_2001a}. We consider an initial velocity field defined as $u = 0.5 (\tanh y)$, and trigger the roll-up of the shear layer with a small $2D$ perturbation. The characteristic velocity and length scales are $\Delta u= u_\textrm{max} - u_\textrm{min}$ and $\delta = \Delta u/(du/dy)_\textrm{max}$, respectively. The Reynolds number is fixed at $Re=\delta \Delta u/\nu=50$ (being $\nu$ the total kinematic viscosity) and the Weissenberg number $Wi=\lambda \Delta u /\delta=25$; moreover, the extensibility $L^2$ is set to $100$, and the ratio of polymeric and solvent viscosity to $0.1$. The $2D$ numerical domain has size $30 \delta \times 100 \delta$, discretised by $128 \times 384$ grid points. Note that, the flow configuration and domain are the same used by \citet{min_yoo_choi_2001a}. \figrefC[a-c]{fig:val-pol} show instantaneous vorticity contours for the flow, where we can observe that the initial perturbation grows in time and generate two vortices (panel a - $t \approx 20 \delta / \Delta u$), which roll-up (panel b - $t \approx 60 \delta / \Delta u$) and eventually merge into one large vortex (panel c - $t \approx 100 \delta / \Delta u$). The quantitative validation is shown in the bottom panel, where we plot the time history of $C_{xx}$ in the center of the domain: symbols are used for the literature results, whereas the red line indicates  our numerical data. As shown by \citet{min_yoo_choi_2001a}, different numerical schemes provide slightly different amplitudes of the maxima of $C_{11}$, with lower values for highly dissipating schemes or with the inclusion of artificial numerical dissipation; in our simulation, we use the high-order WENO scheme for the advection of the conformation tensor and neglect any artificial dissipation, and we find good agreement of the deformation time history over the whole vortex merging process \citep{rosti_brandt_2017a, de-vita_rosti_izbassarov_duffo_tammisola_hormozi_brandt_2018a, rosti_izbassarov_tammisola_hormozi_brandt_2018a, izbassarov_rosti_niazi-ardekani_sarabian_hormozi_brandt_tammisola_2018a}.

\begin{figure}
\centering
\includegraphics[width = 0.325\textwidth]{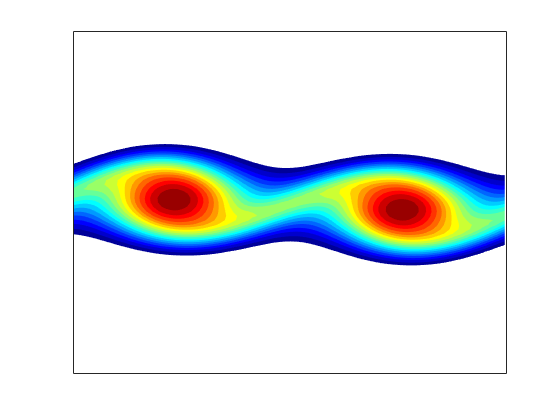}
\includegraphics[width = 0.325\textwidth]{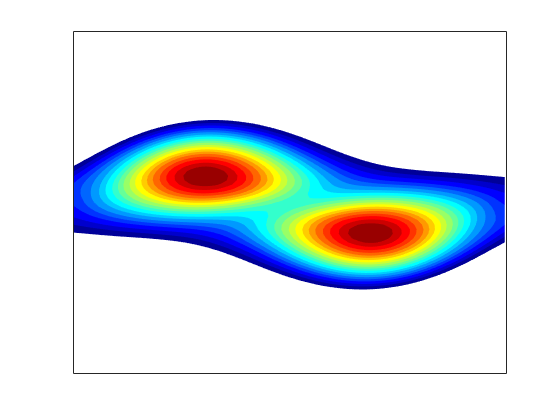}
\includegraphics[width = 0.325\textwidth]{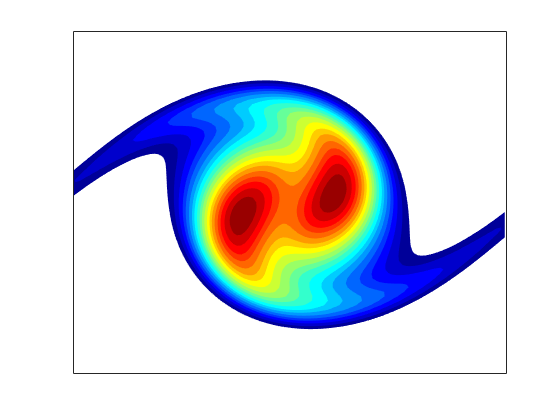}
\includegraphics[width = 0.49\textwidth]{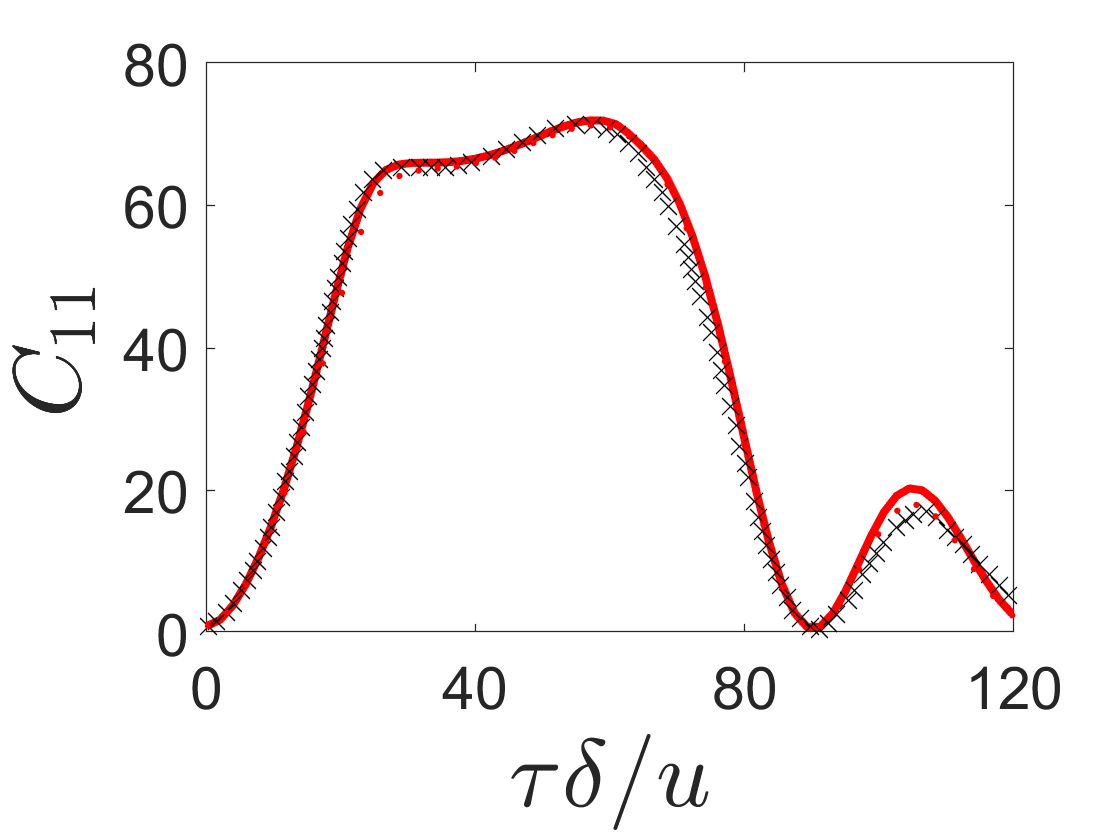}
\caption{(a-c) Instantaneous contours of the absolute value of vorticity at time $t \approx 20$, $60$ and $100 \delta / \Delta u$. The color scale from blue to red ranges from $0.05$ to $0.4$ in (a), from $0.05$ to $0.3$ in (b) and from $0.05 $ to $0.25$ in (c). (d) Time evolution of the  $C_{11}$ component of the polymer conformation tensor. The red line displays our numerical results, while the symbols those by \citet{min_yoo_choi_2001a}.}
\label{fig:val-pol}
\end{figure}

\section{Results} \label{sec:result}
We study turbulent duct flows with polymers and start our analysis by considering two cases at $Re=2800$ (based on the bulk velocity $U_b$, duct half height $h$ and total kinematic viscosity $\nu$),  Newtonian fluid and the viscoelastic fluid consisting of a suspension of polymers as introduced in the previous section. The chosen polymer solution is characterized by the following parameters: $\beta=0.9$, $L^2=3600$ and $Wi=1.5$ ($Wi_{\tau_0}=\lambda u^2_{\tau_0} /\nu = 18$ being $u_{\tau_0}$ the friction velocity of the Newtonian simulation). Using the previous set of parameters, the Newtonian case has an average friction Reynolds number $Re_\tau$ equal to $183$, while the polymer one provides a value of $155$ which corresponds to a drag reduction $DR$ of approximately $29\%$, computed as $DR=-\Delta \tau_w/\tau_{w0}$, where $\Delta \tau_w = \tau_w - \tau_{w0}$, being $\tau_w$ the total shear stress at the wall (as in the computation of the friction velocity), averaged over the entire four sides, and $\tau_{w0}$ the value for the Newtonian case. The Newtonian results will be represented hereafter in blue and the polymer ones in red.

\subsection{Flow statistics}
\begin{figure}
\centering
\includegraphics[height = 0.35\textwidth]{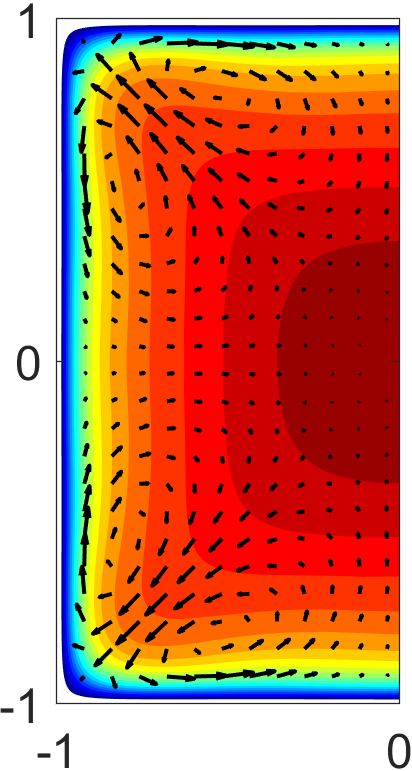}
\includegraphics[height = 0.35\textwidth]{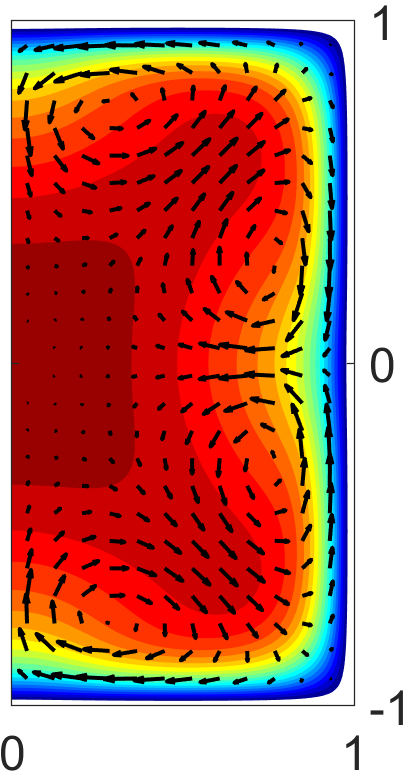}
\includegraphics[height = 0.35\textwidth]{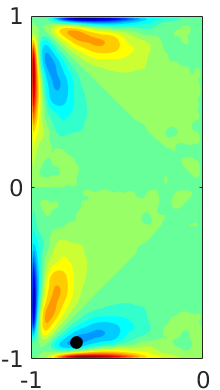}
\includegraphics[height = 0.35\textwidth]{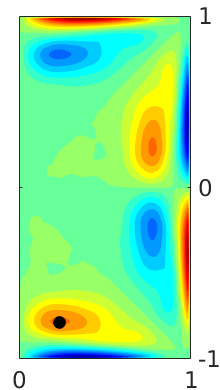}
\caption{Contour of the mean streamwise component of (left) velocity $\overline{u}$ and (right) vorticity $\overline{\omega}_x$. In each figure, the left and right half of the duct are used for the Newtonian and viscoelastic fluids, respectively. The color scale ranges from $0$ (blue) to $1.4U_b$ (red) in the left panel and from $-0.5U_b/h$ (blue) to $0.5U_b/h$ (red) in the right one. The two left panels also report the in-plane mean velocity with arrows.}
\label{fig:mean}
\end{figure}
\begin{figure}
\centering
\includegraphics[height = 0.35\textwidth]{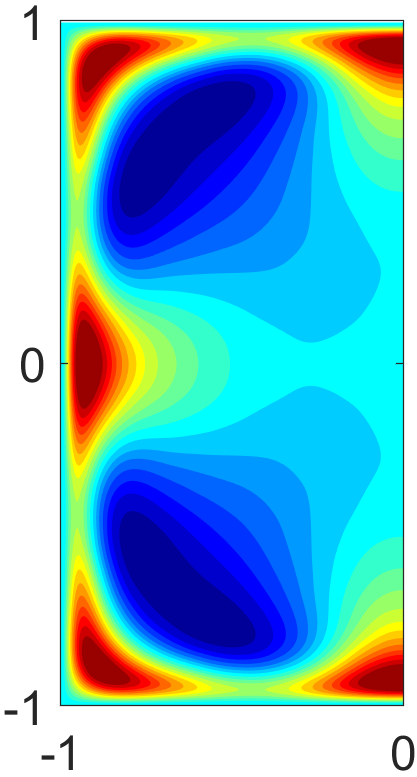}
\includegraphics[height = 0.35\textwidth]{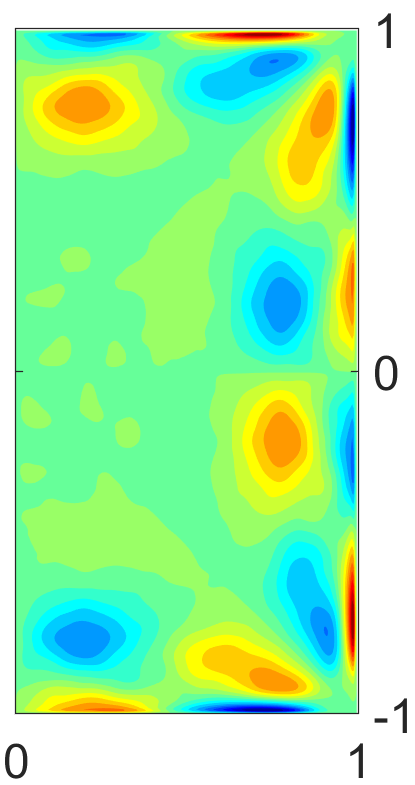}
\caption{Contour of the difference of the mean streamwise component of the velocity $\overline{u}$ (left)  and vorticity $\overline{\omega}_x$ (right) between the Newtonian and viscoelastic fluid. The color scale ranges from $-0.2$ (blue) to $0.2U_b$ (red) in the left panel and from $-0.5U_b/h$ (blue) to $0.5U_b/h$ (red) in the right one.}
\label{fig:meandiff}
\end{figure}

\figrefAC{fig:mean} shows the contour of the mean streamwise velocity component $\overline{u}$ (panel a) and of the mean streamwise vorticity component $\overline{\omega}_x$ (panel b), where in each panel, the left half of the duct reports the data for the Newtonian flow, whereas the right side the viscoelastic counterpart. For major clarity, \figrefA{fig:meandiff} shows the difference between the Newtonian and viscoelastic flow. The Newtonian case shows the streamwise velocity contour typical of a duct flow, characterized by the symmetry inherited from the considered geometry. The maximum streamwise velocity is located at the center of the duct and equals $1.3U_b$. The streamwise vorticity displays the in-plane secondary flow typical of turbulent duct flows: $8$ regions ($4$ in the half section shown in the figure) can be easily identified with alternate sign of vorticity, located symmetrically with respect to the horizontal and vertical lines passing through the center and with the two diagonals. In particular, we mark with a filled black circle the location of one of the maxima of the vorticity, which we find at $y=-0.74h$, $z=-0.9h$ in the Newtonian case. In the polymeric flows, the streamwise velocity and vorticity components are altered (see \figrefA{fig:meandiff}): indeed, the two flows differ mainly in the near wall region, with strong differences in the streamwise velocity close to the corners. The difference in the streamwise vorticity component is mostly concentrated along the edges. In particular, the streamwise velocity contour exhibits higher values close to the corners and, at the same time, the secondary flow is modified, with the locations of the maximum vorticity moving towards the center, \ie they are displaced away from the walls. In fact, the coordinates of the selected maximum vorticity become $y=0.23h$, $z=0.78h$. The magnitude of the vorticity maxima are also modified, reducing by $20\%$ from $0.29U_b/h$ in the Newtonian fluid to $0.23U_b/h$ in the presence of polymers, while the overall integral of the vorticity in each of the $8$ sectors increases by $17\%$ from $0.023U_b/h$ in the Newtonian flow to $0.027U_b/h$ in the viscoelastic one. Thus, although the peak value of the in-plane vorticity is reduced, overall it is enhanced by the polymer addition due to an enlargement of the area with non-zero vorticity. The modification of the secondary flow is further analysed by the maximum of the magnitude of $\overline{v}$ in the domain; indeed, this slightly decreases from approximately $0.019U_b$ in the Newtonian flow to $0.018U_b$ in the polymer solution.

\begin{figure}
\centering
\includegraphics[width = 0.325\textwidth]{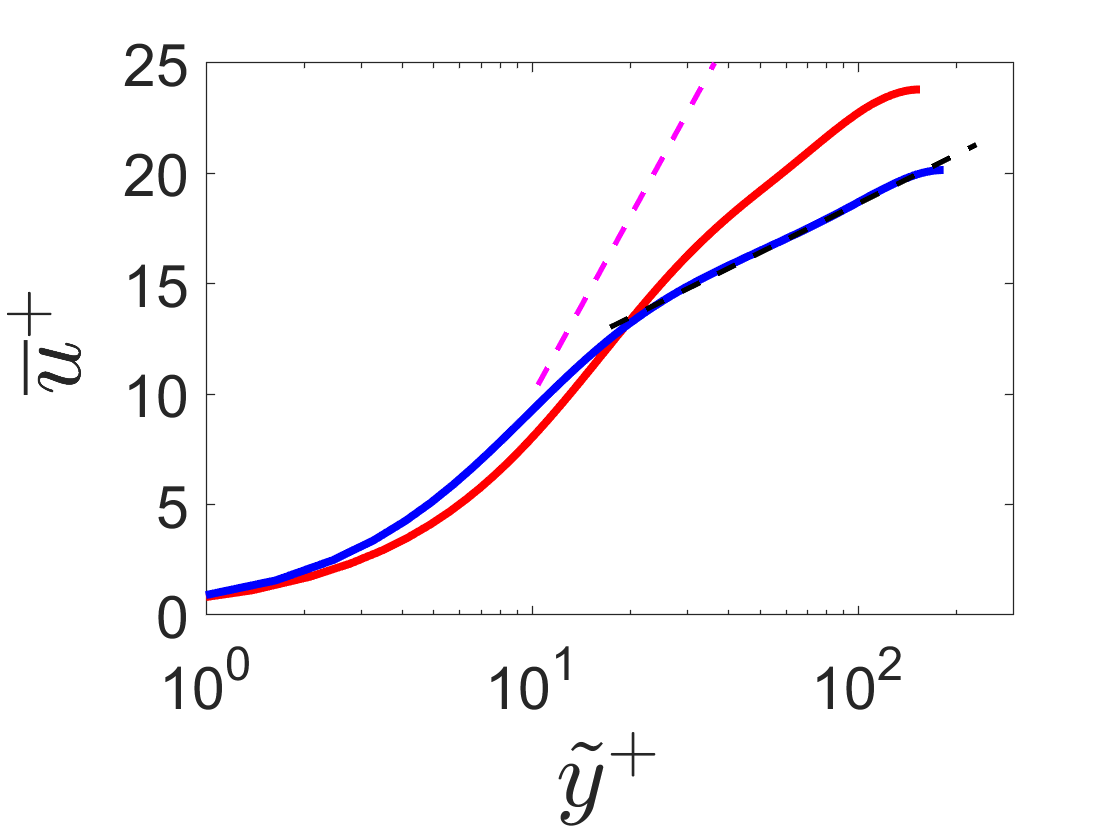}
\includegraphics[width = 0.325\textwidth]{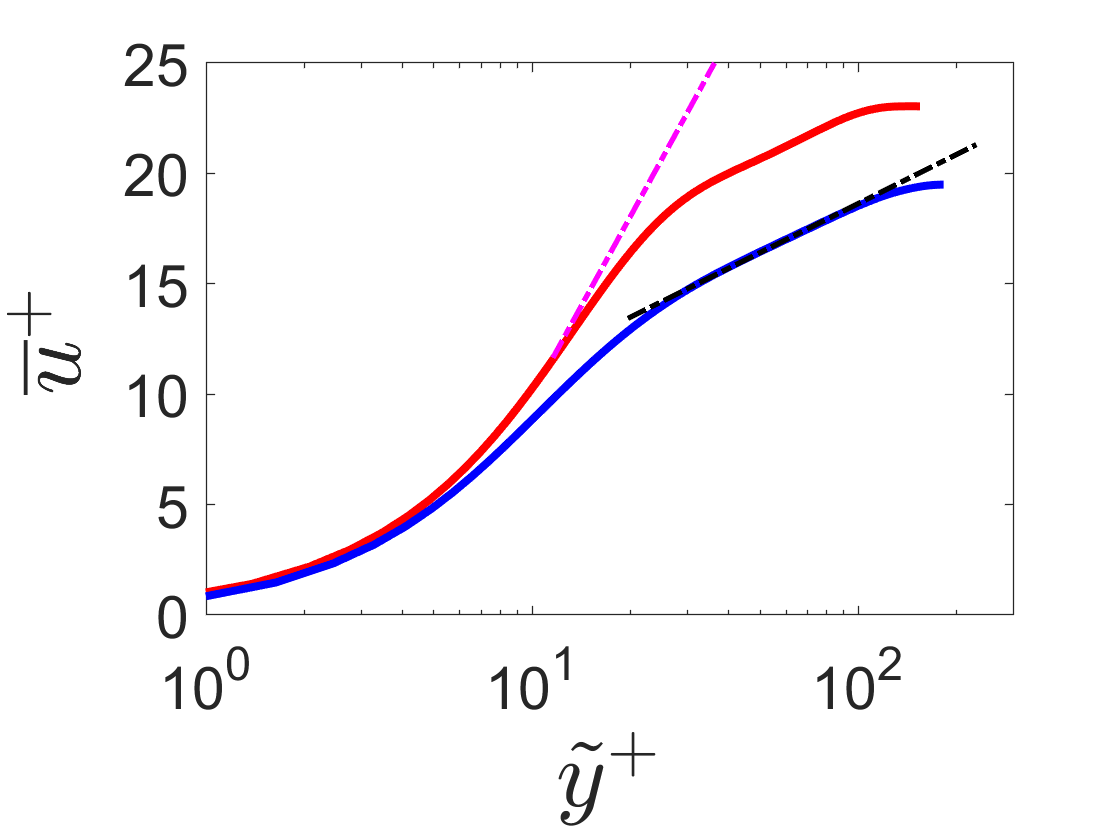}
\includegraphics[width = 0.325\textwidth]{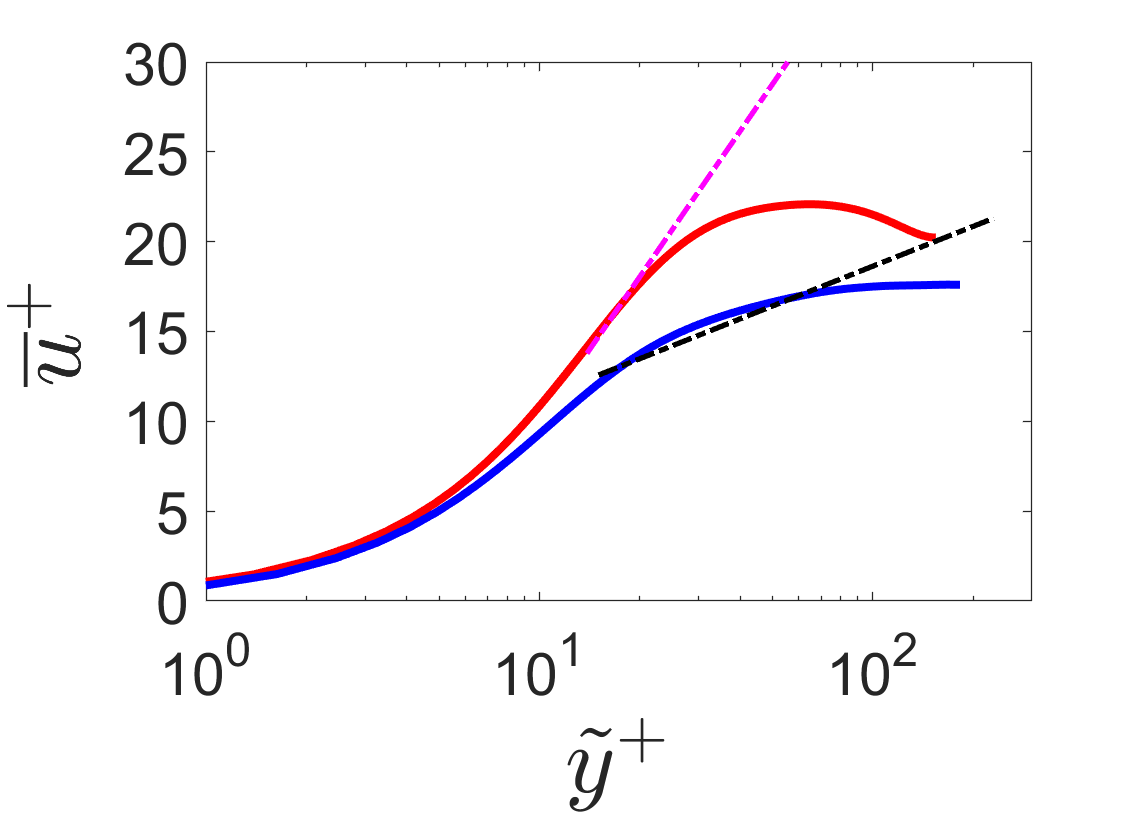}
\caption{Mean streamwise velocity as function of the wall-normal distance in logarithmic scale and wall units. The blue and red lines are used for the numerical results without and with polymers. The magenta dash dotted line is the polymer Maximum Drag Reduction \citep{virk_1975a, lvov_pomyalov_procaccia_tiberkevich_2004a} while the black one the Newtonian log law. The three panels correspond to different sections: (left) $z=0$, (middle) $z=0.3h$, and (right) $z=0.6h$.}
\label{fig:loglaw}
\end{figure}
\figrefAC{fig:loglaw} shows the mean streamwise velocity component $\overline{u}$ as a function of the wall normal distance $\widetilde{y}$ (both in wall units). The left panel shows the mean velocity profile in the middle plane ($z=0$), the middle panel the velocity at $z=0.3h$, and the right one at $z=0.6h$. In the middle plane, where the secondary flow is weak, we can identify three regions in the velocity profile of the Newtonian fluid (left panel) similarly to what is found for a turbulent channel flow: first, the viscous sublayer for $\widetilde{y}^{+}<5$ where the variation of $\overline{u}^{+}$ with $\widetilde{y}^{+}$ is approximately linear; then, the so-called log-law region, $\widetilde{y}^{+}>30$, where the variation of $\overline{u}^{+}$ versus $\widetilde{y}^{+}$ is logarithmic; finally, the region between $5$ and $30$ wall units is called the buffer layer and neither laws hold. As we approach the side walls, $x=0.3h$ and $0.6h$, the range of the equilibrium layer where the velocity profile is logarithmic reduces (middle panel), eventually disappearing (right panel). The profiles have similar trends for the viscoelastic fluid, in which case, however, the inertial range has an upward shift, consistent with the observed drag reduction \citep{warholic_massah_hanratty_1999a}. Note that, all the data stays below the MDR curve \citep{virk_1975a, lvov_pomyalov_procaccia_tiberkevich_2004a}, a limit curve approached only in the case of very high $Wi$, shown as a dashed line in the figure.

\begin{figure}
\centering
\includegraphics[width = 0.325\textwidth]{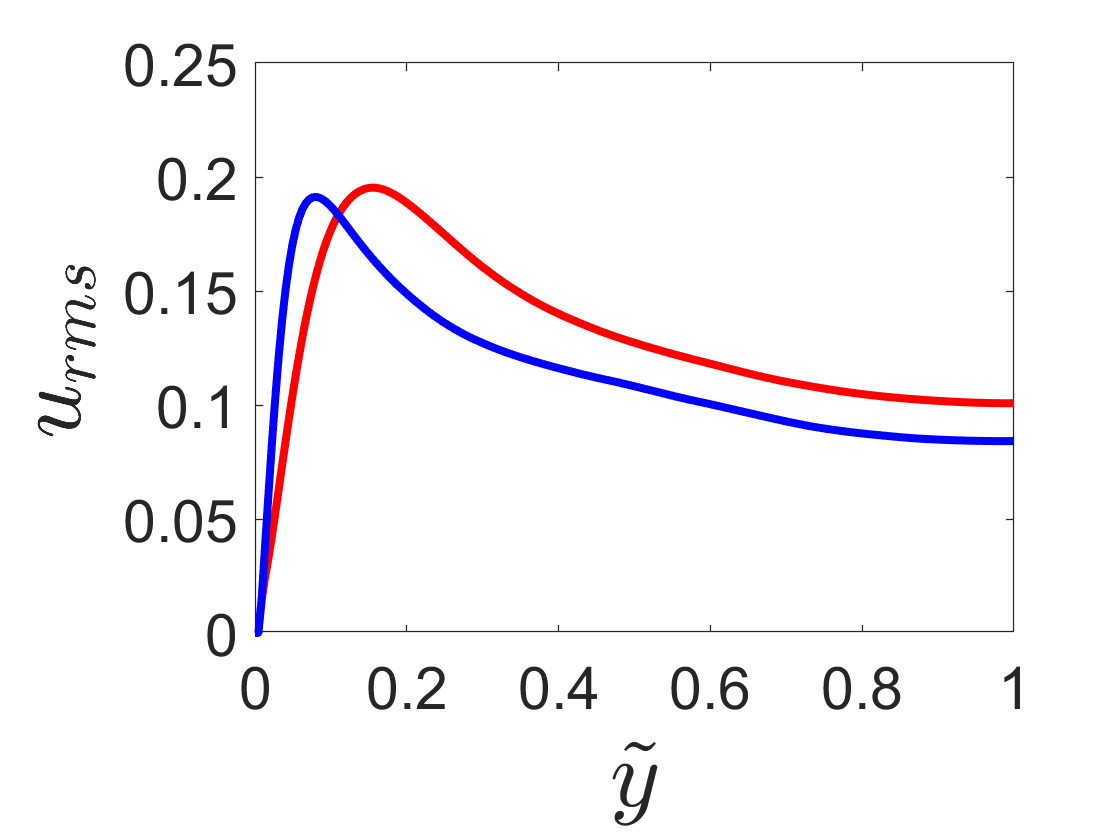}
\includegraphics[width = 0.325\textwidth]{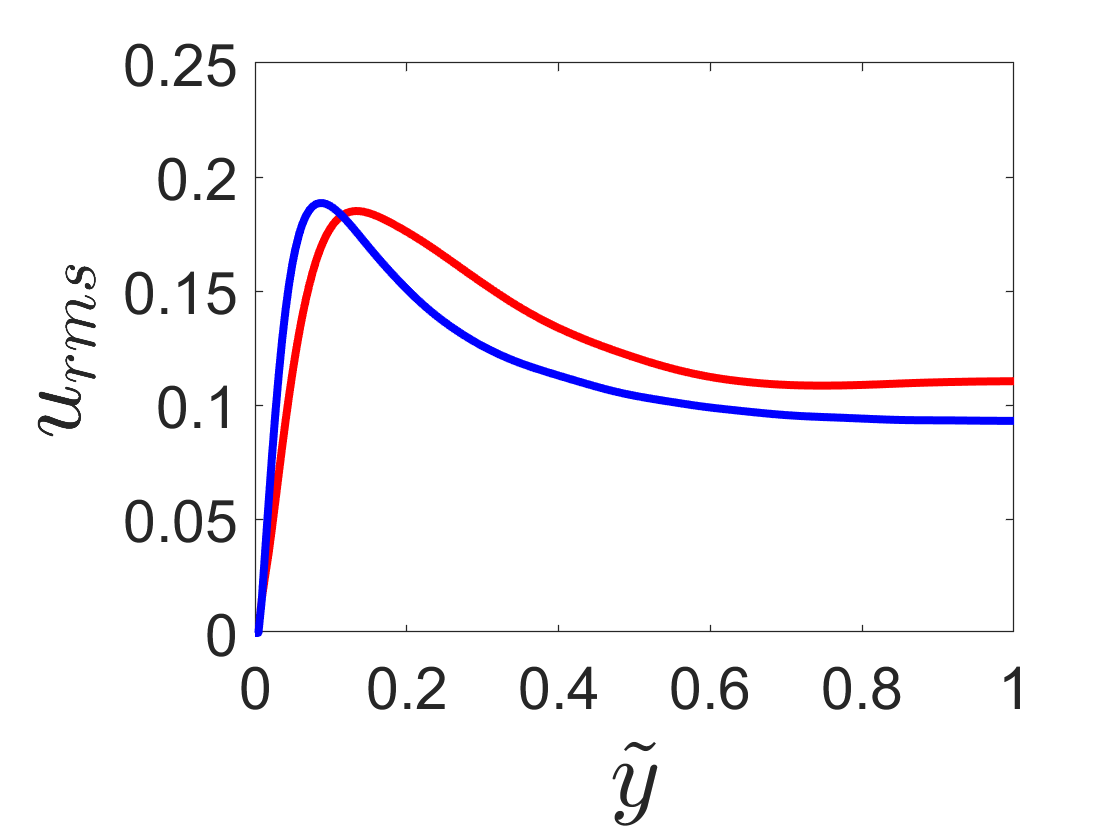}
\includegraphics[width = 0.325\textwidth]{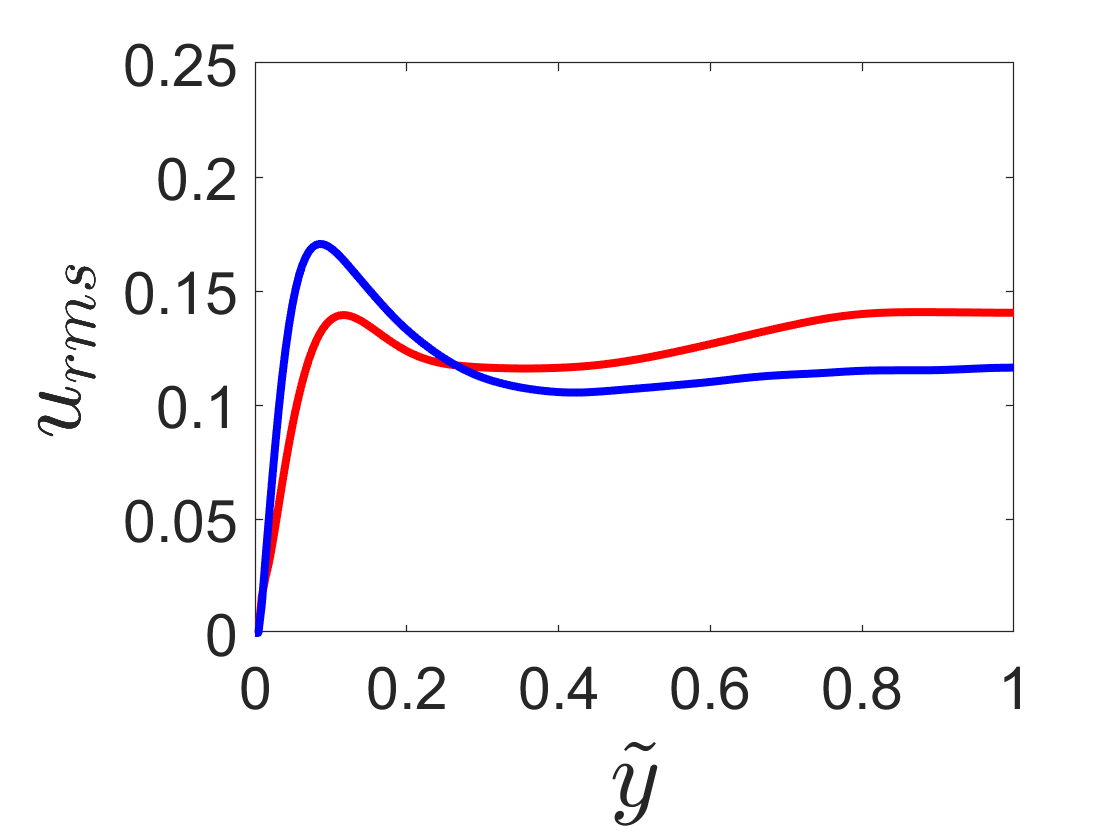}
\includegraphics[width = 0.325\textwidth]{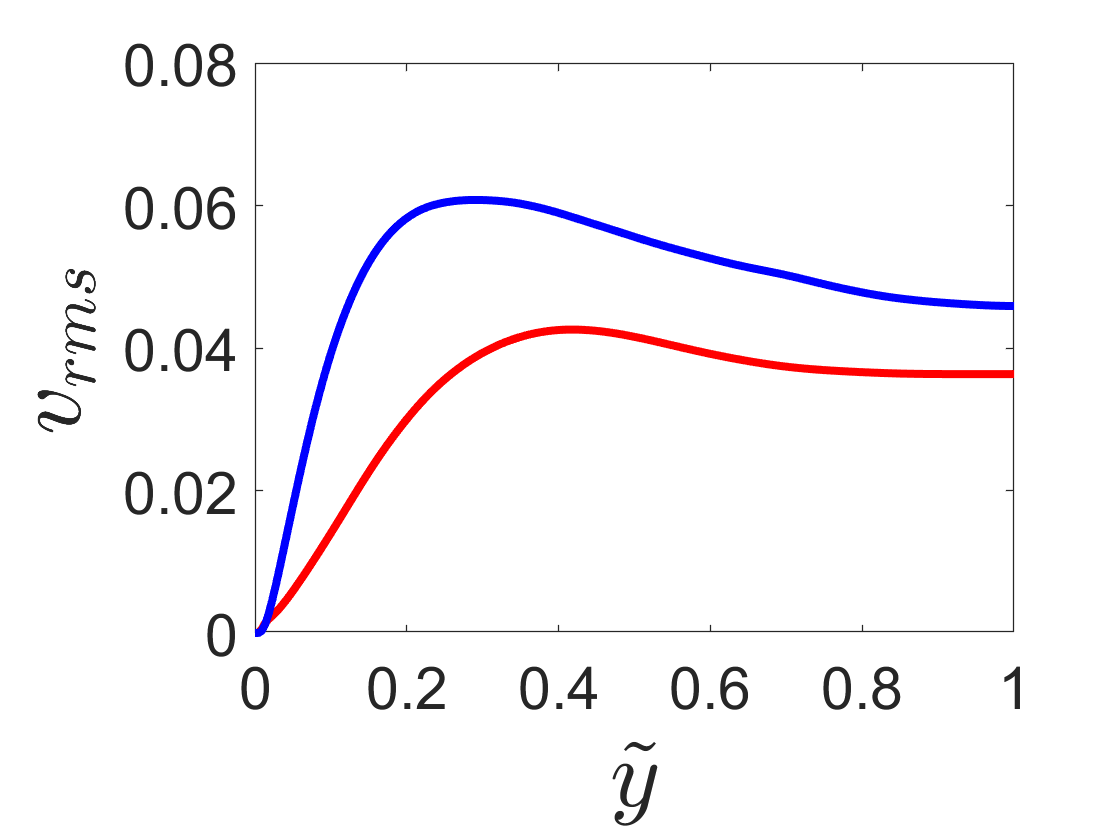}
\includegraphics[width = 0.325\textwidth]{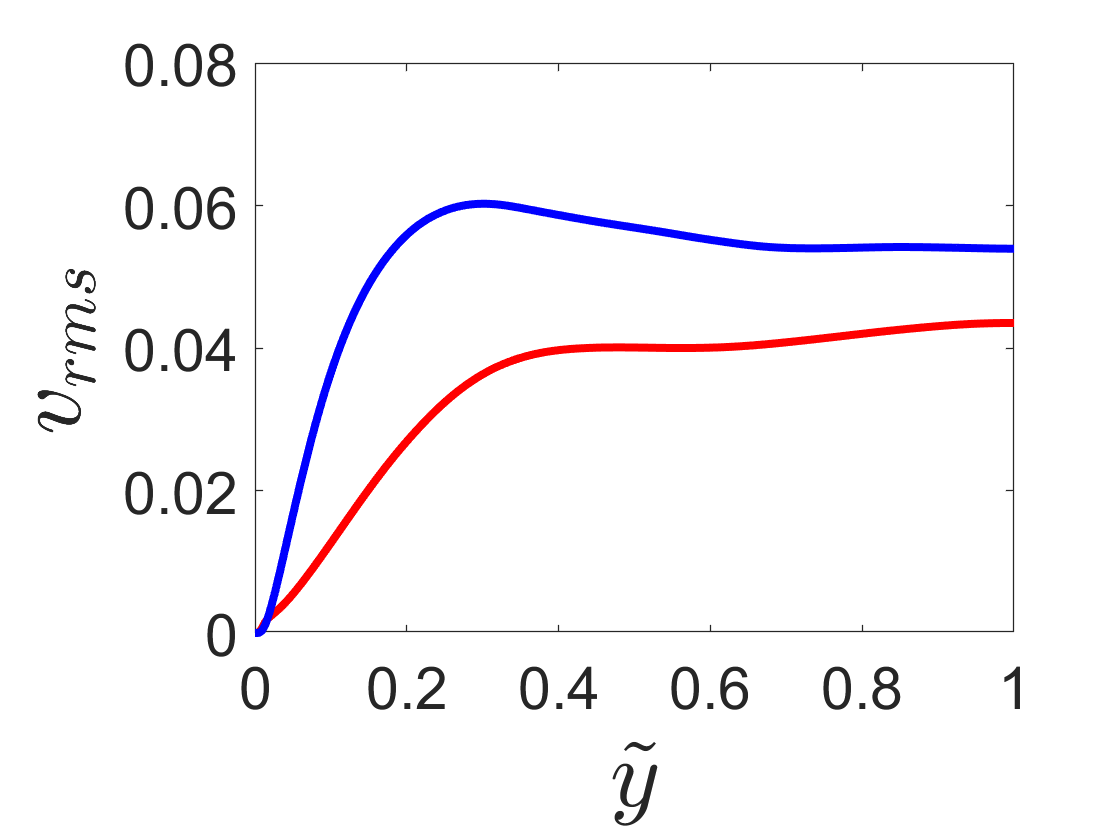}
\includegraphics[width = 0.325\textwidth]{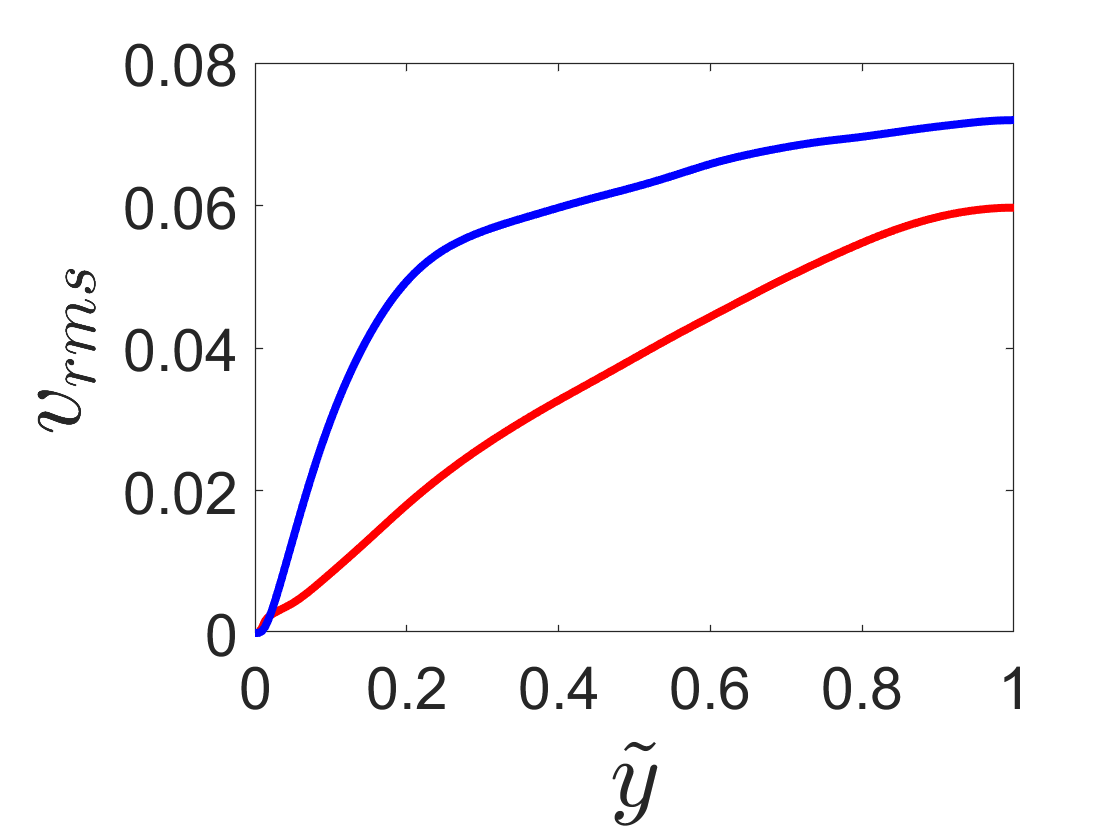}
\includegraphics[width = 0.325\textwidth]{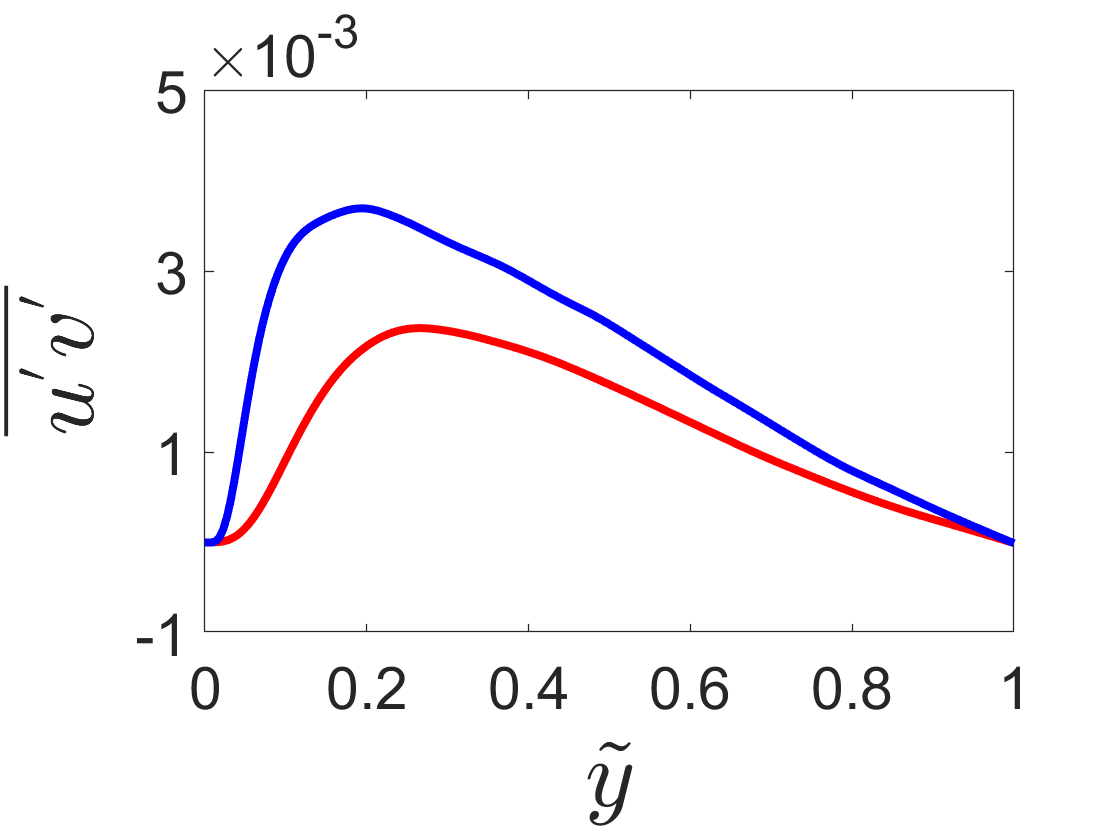}
\includegraphics[width = 0.325\textwidth]{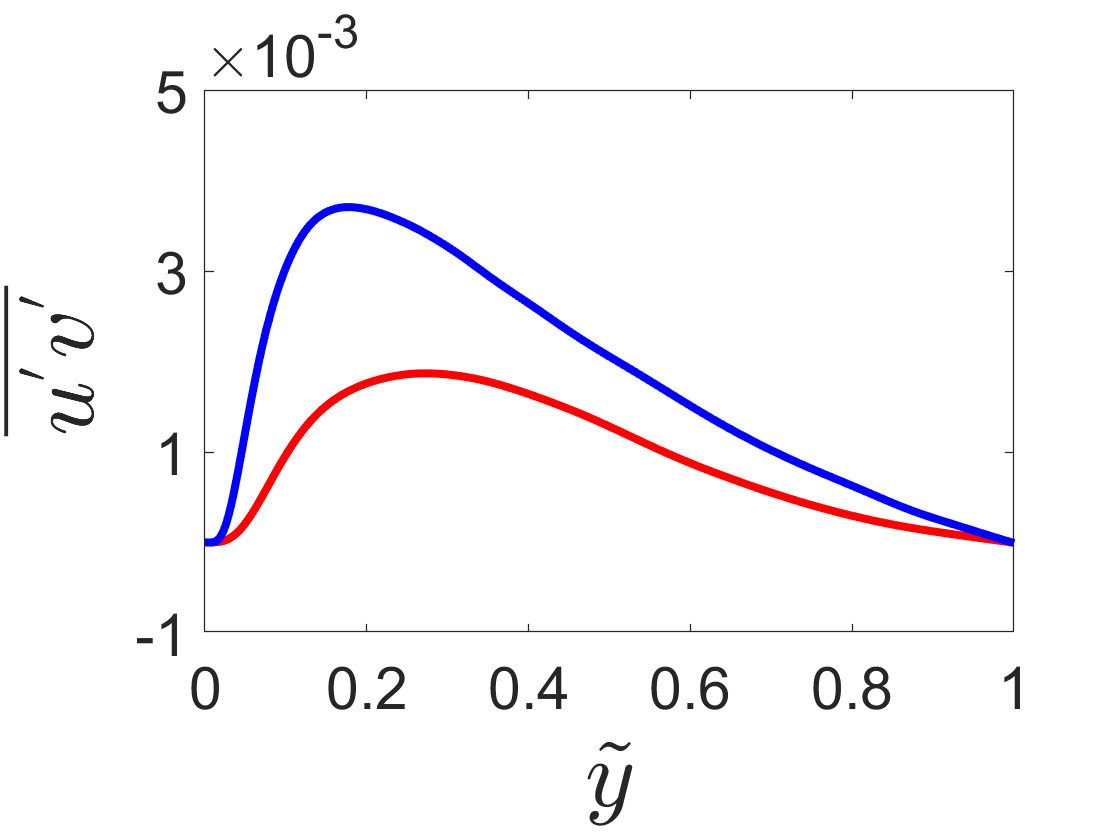}
\includegraphics[width = 0.325\textwidth]{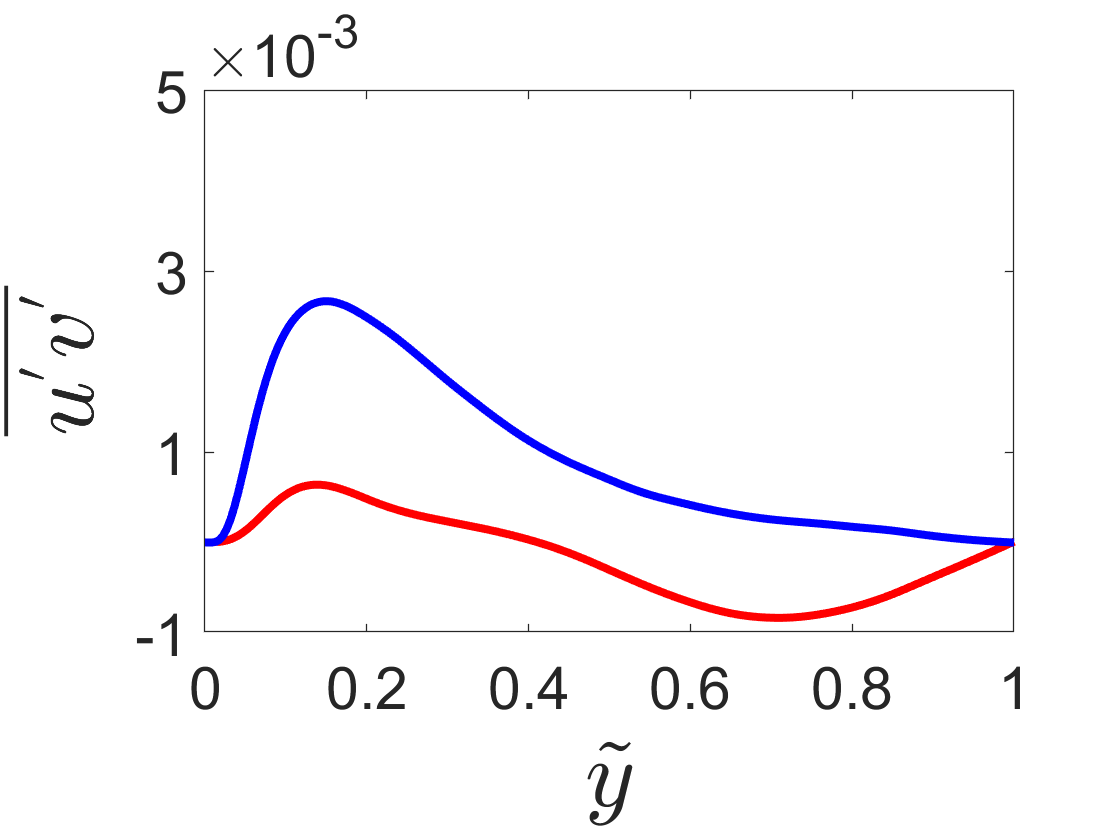}
\caption{Wall-normal mean profile of the (top row) streamwise rms velocity fluctuation, (middle row) wall-normal rms velocity fluctuation, and (bottom row) cross term of the Reynolds stress tensor $u'v'$. The three columns correspond to different sections: (left) $z=0$, (middle) $z=0.3h$, and (right) $z=0.6h$. The line and color style is the same as in \figrefA{fig:loglaw}.}
\label{fig:rey}
\end{figure}
\figrefAC{fig:rey} displays the velocity fluctuations profile as a function of the wall-normal distance $\widetilde{y}$. In particular, the top row reports the root mean square (rms) of the streamwise component of the velocity fluctuations $u_{rms}$, the middle panel the rms of one of the wall normal components $v_{rms}$, and the bottom row the cross term of the Reynolds stress tensor, $u'v'$. The Newtonian case shows a peak of $u_{rms}$ close to the wall, with slightly different values in the three sections, decreasing when moving from the middle plane towards the side-walls. Away from the wall, the streamwise velocity fluctuations reach a plateau with a local minimum whose amplitude is approximately half the peak value in the centerline. On the other hand, the wall-normal velocity fluctuations have a weak maximum close to the wall in the middle line, which then decreases at $z=0.3h$, and eventually disappears in $z=0.6h$, when approaching the corners. The flow with polymers displays similar r.m.s. profiles; however, the peaks of both $u_{rms}$ and $v_{rms}$ are located further away from the wall; also, a strong decrease in the wall-normal fluctuation $v_{rms}$ is observed in all sections, whereas an increase in the streamwise component is evident only close to the centerline, as observed in plane channels. Finally, we consider the cross term of the Reynolds stress tensor $u'v'$; the profiles show a peak close to the wall and a value approaching $0$ towards the centerline due to symmetry. The profile at $z=0.3h$ has a lower peak, and still decreases monotonically after its maximum value; on the other hand, the Reynolds stress $u'v'$ closer to the side walls, $z=0.6h$, goes to zero rapidly and remains approximately zero until $\tilde{y}=h$. In the cases with polymers, the Reynolds shear stress is much weaker than in the Newtonian cases; it even becomes negative away from the wall in the section at $z=0.6h$. This is due to the transverse shear near the walls leading to additional vorticity (\figrefA{fig:mean}); this effect is more pronounced in the viscoelastic case than in the Newtonian flow, due to the enhancement of the near wall vorticity, as shown in the right panel of \figrefA{fig:meandiff}. Overall, the velocity fluctuation data indicate that the polymer solution has enhanced streamwise velocity fluctuations, and strongly reduced in-plane components and shear stresses, in good agreement with many previous studies in the literature \cite[see \eg the collection of data from numerous authors in][]{escudier_nickson_poole_2009a}.

\begin{figure}
\centering
\includegraphics[width = 0.45\textwidth]{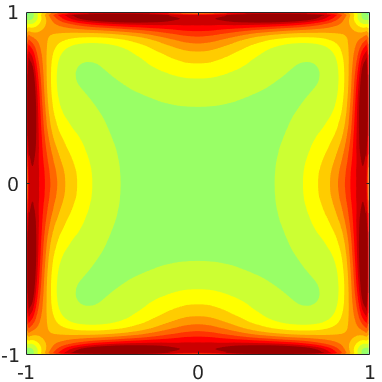}
\includegraphics[width = 0.45\textwidth]{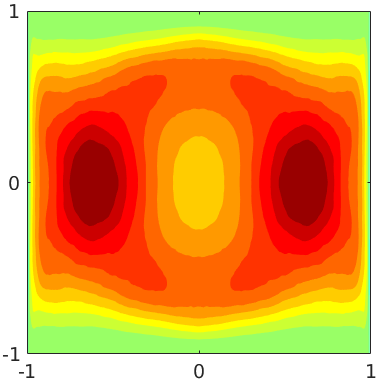}\\ \vspace{0.65cm}
\includegraphics[width = 0.45\textwidth]{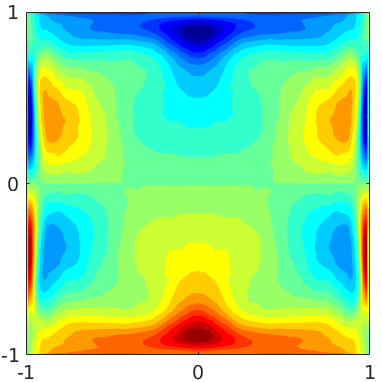}
\includegraphics[width = 0.45\textwidth]{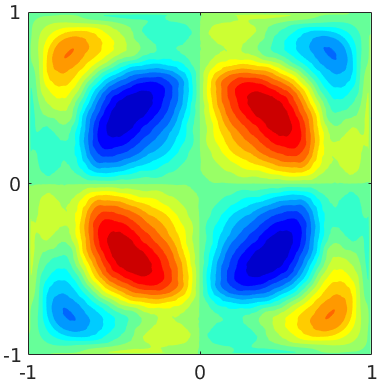}
\caption{Contours of four components of the conformation tensor $C_{ij}$: (top left) $C_{11}$, (top right) $C_{22}$, (bottom left) $C_{12}$, and (bottom right) $C_{23}$. Color scales range from $30$ (green) to $800$ (red), from $5$ (blue) to $50$ (red), from $-30$ (blue) to $30$ (red), and from $-10$ (blue) to $10$ (red), respectively.}
\label{fig:cij}
\end{figure}
We conclude the analysis of the main flow statistics by showing in \figrefA{fig:cij} four components of the conformation tensor $C_{ij}$. Note that, the conformation tensor is strongly related to the added stress term due to the polymers (see \equref{eq:polymerStress}). In particular, in the figure we show the streamwise diagonal component $C_{11}$, one of the two diagonal in plane components $C_{22}$, the corresponding off-diagonal term $C_{12}$, and the in plane off-diagonal term $C_{23}$. The other components can be deduced from symmetry arguments: in particular $C_{33}$ and $C_{13}$ can be obtained from $C_{22}$ and $C_{12}$ by symmetry around the bisection of the quadrants I and III. As expected, the streamwise component is the dominant one, with strong peak values located close to the wall, while interestingly, in the four corners and in the center of the duct $C_{11}$ reaches its minimum value. Also, $C_{11}$ is the only diagonal component of the conformation tensor showing the full $8$ symmetries; indeed, $C_{22}$ presents only the horizontal and vertical symmetries, with the diagonal ones recovered through the combination with the other component $C_{33}$ (not shown here). In particular, the in plane diagonal component $C_{22}$ is minimum at the wall, and acts mainly on the horizontal centerline. The cross terms $C_{12}$ has its peak located at the walls, while the component $C_{23}$ has its peaks on the two diagonals. The in plane shear component $C_{23}$ is not zero close to the corners, and over the two diagonals, where it shows a series of alternating maxima and minima, in a similar fashion to the shear Reynolds stress which exhibits alternate signs moving from one wall to the other (\figrefA{fig:rey}).
\begin{figure}
\centering
\includegraphics[width = 0.325\textwidth]{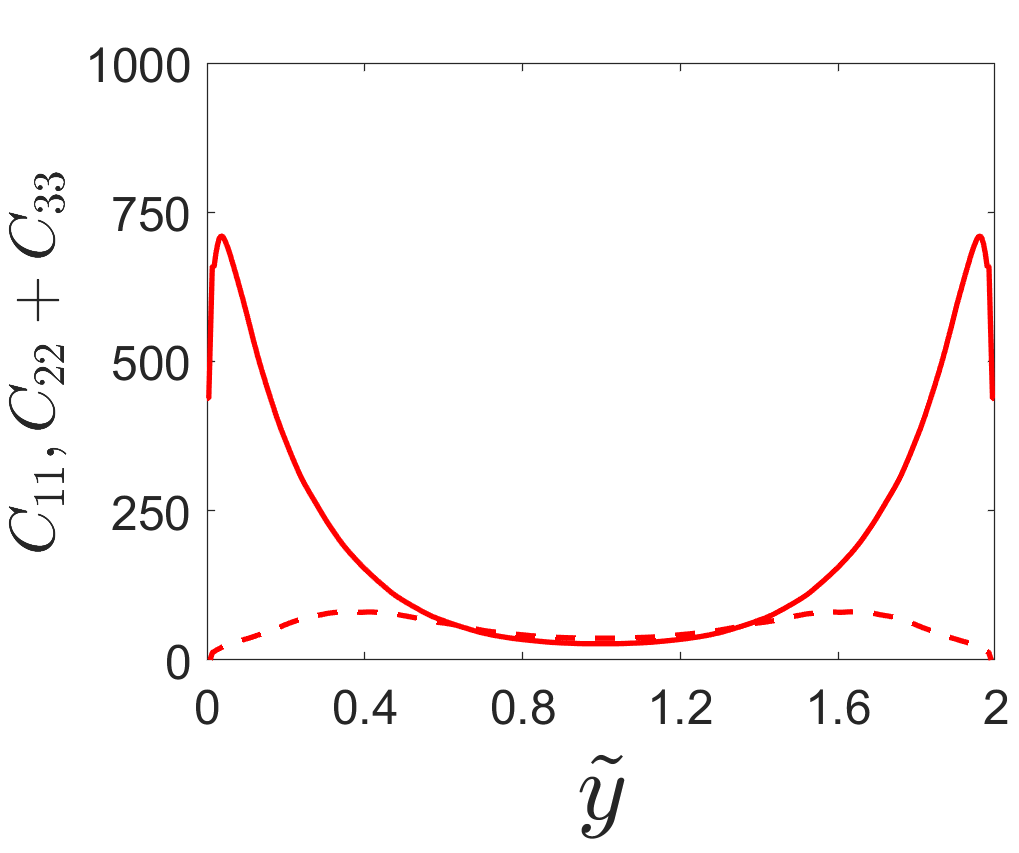}
\includegraphics[width = 0.32\textwidth]{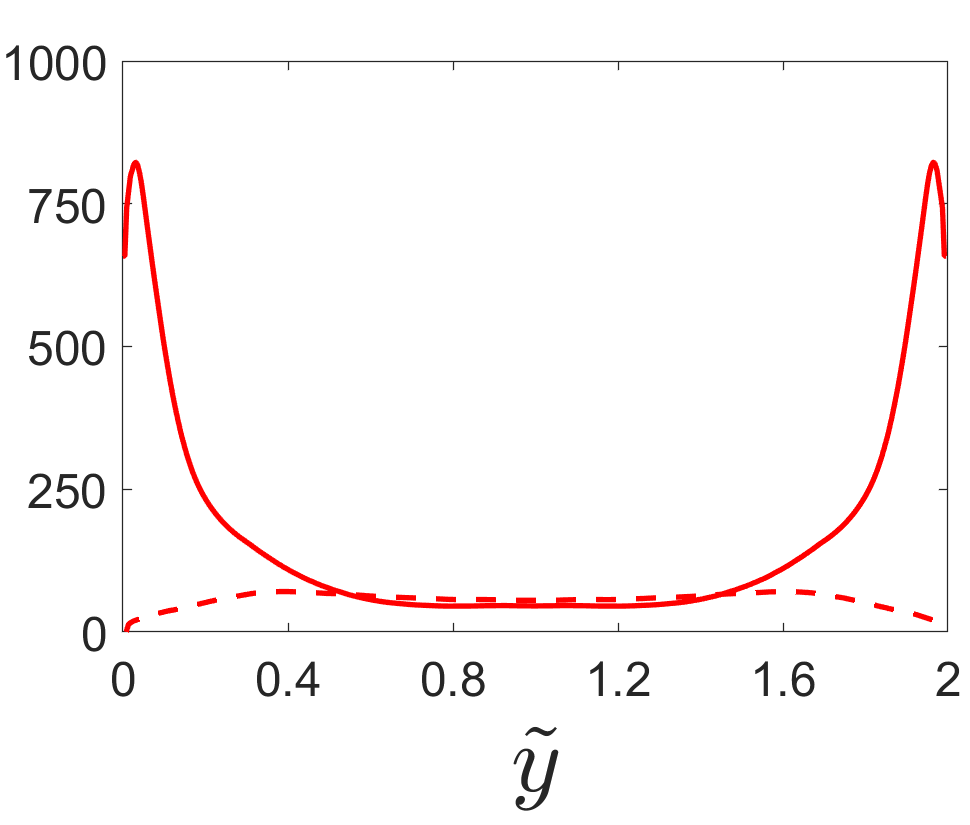}
\includegraphics[width = 0.32\textwidth]{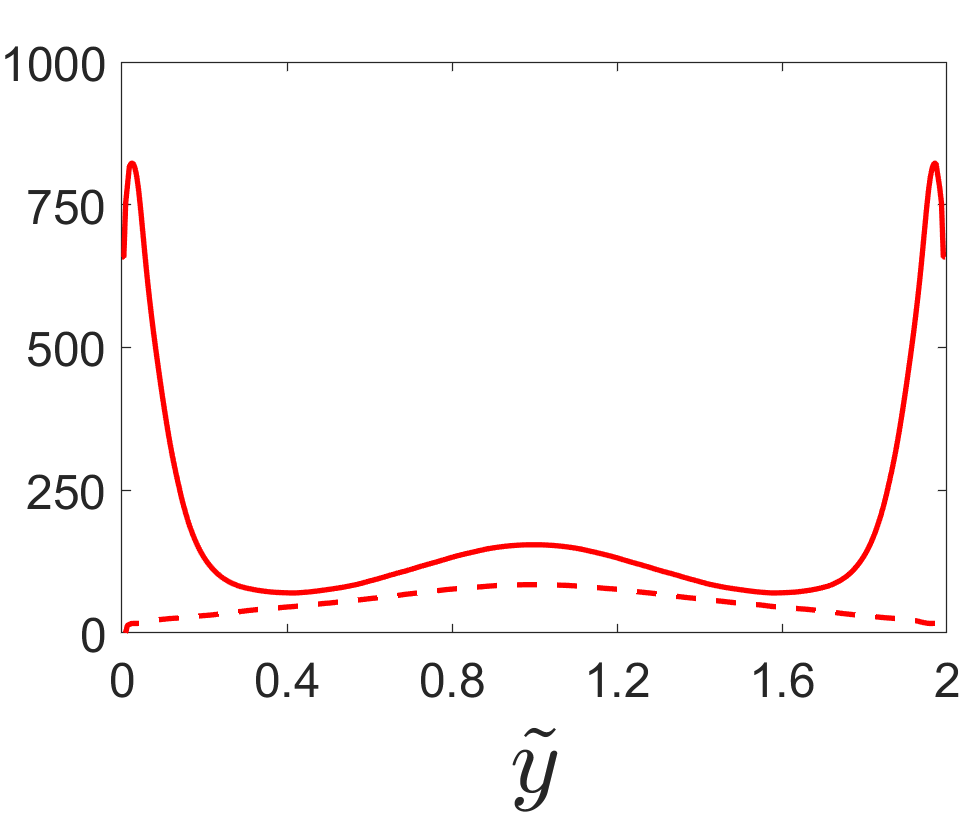}
\caption{Mean profile of the streamwise $C_{11}$ (solid line) and in-plane component $C_{22}+C{33}$ (dashed line) of the conformation tensor $C_{ij}$. The three panels correspond to different sections: (left) $z=0$, (middle) $z=0.3h$, and (right) $z=0.6h$.}
\label{fig:cijp}
\end{figure}
\figrefAC{fig:cijp} shows the wall-normal profiles of $C_{11}$ (solid line) and $C_{22}+C_{33}$ (dashed line) in the middle plane of the duct ($z=0$), at $z=0.3h$ and at $z=0.6h$ in the left, middle and right panels, respectively. Note that, $C_{22}+C_{33}$ is a measure of the in-plane stretching of the polymers, while its sum with $C_{11}$ is the trace of the conformation tensor. We observe that $C_{11}$ (and the trace) has a peak close to the wall and then decreases to its minimum value in the center of the duct. Also, the peak value increases moving from the middle plane towards the side walls; $C_{11}$ starts increasing for all $\tilde{y}$ close to the side walls ($z=0.6h$) as we are approaching the side walls peaks (see the top left panel in \figrefAC{fig:cij}). The in-plane stretching ($C_{22}+C_{33}$) is more moderate than the streamwise one, and the profiles  show minimum values at the wall. Also, the peak values of $C_{22}+C_{33}$ are located around $h/2$ in the middle plane and move towards $\tilde{y}=1$ when approaching the side walls (see the top right panel in \figrefAC{fig:cij}). \figrefAC{fig:cij} and \figrefAC{fig:cijp} show that the polymers mainly elongate in the streamwise direction in the near wall region and away from the corners; however, in-plane elongation originating from the secondary cross flow is present as well, especially at a distance from the walls of approximatively $h/2$.

\begin{figure}
\centering
\includegraphics[height = 0.42\textwidth]{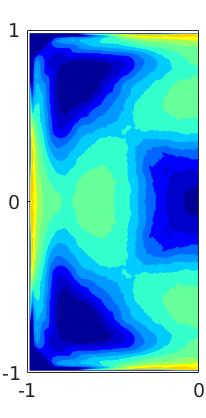}
\includegraphics[height = 0.42\textwidth]{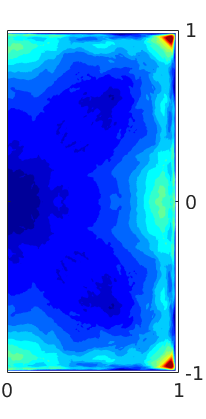}
\includegraphics[height = 0.42\textwidth]{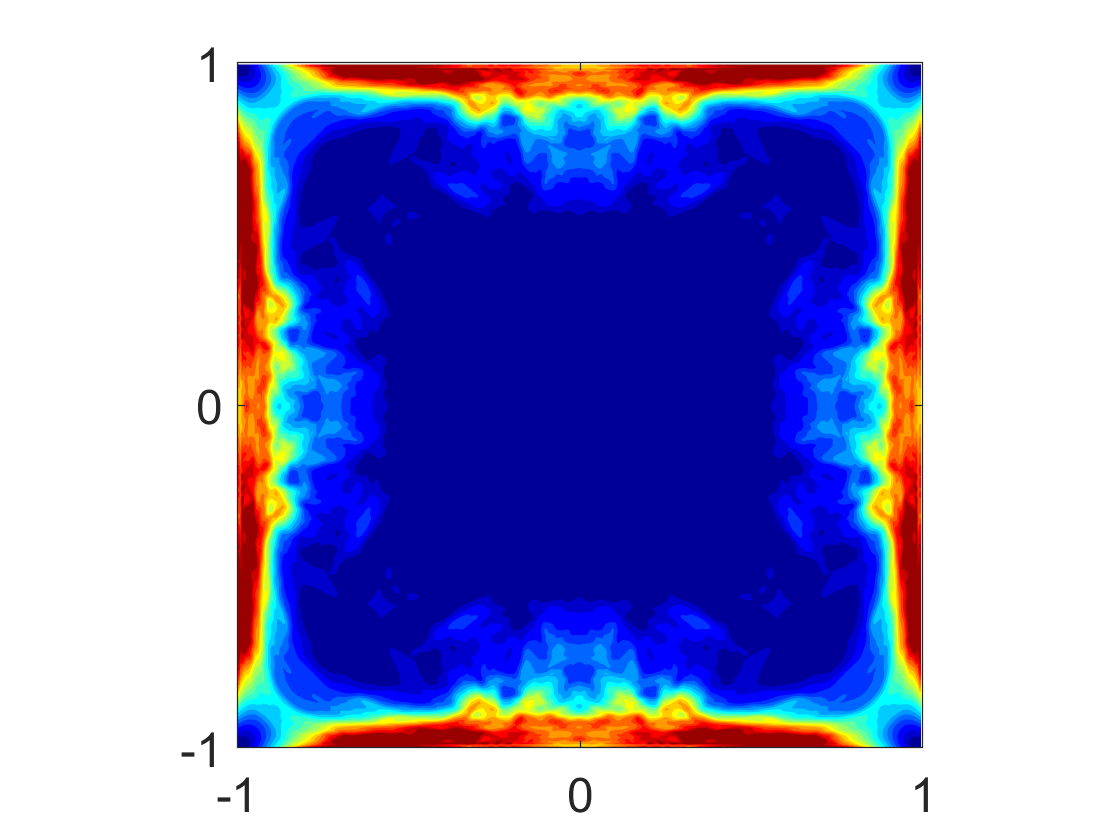}
\caption{(left) Cross-correlations of the streamwise (left half) and wall-normal (right half) velocities and polymer force, defined in \equref{eq:crosscorr}. Color scales range from $-1$ to $1$, and from $-0.25$ to $0.25$, respectively. (right) Contour of the mean first normal difference $\mathcal{N}_1$, with the color scales ranging from $0$ (blue) to $0.35$ (red).}
\label{fig:cijCorr}
\end{figure}

Next, \figrefC[left]{fig:cijCorr} shows the cross-correlation $\rho_i$ defined as
\begin{equation} \label{eq:crosscorr}
\rho_i= \dfrac{\overline{u_i f_i}}{{u_i}_{rms} {f_i}_{rms}},
\end{equation}
where $f_i$ is the polymer volume force, \ie the contribution to the Navier-Stokes equation of the polymeric stress tensor $f_i=(1-\beta)/Re \partial \tau_{ij}/\partial x_j$. The left half of the plot shows the streamwise component $\rho_1$, while the right half the two identical wall-normal ones $\rho_2=\rho_3$.  In the streamwise direction ($\rho_1$), the polymer contribution is correlated with the corresponding velocity component in the near wall region of the mid-plane far from the corners, thus enhancing fluctuations in this particular region. On the contrary, the streamwise polymer body force and velocity become anti-correlated in the bulk of the flow away from the walls, where the polymer contribution opposes the turbulent fluctuations. This is similar to what was found by \citet{dubief_terrapon_white_shaqfeh_moin_lele_2005a} for a turbulent plane channel flow. However, as the corners are approached, the behavior drastically changes and the two streamwise components are strongly anti-correlated everywhere, even in the near wall region, thus indicating that the addition of polymers tends to suppress fluctuation in the corners, as also shown in the right column of figure \ref{fig:rey}. The wall-normal component of the cross-correlation $\rho_{2,3}$ shows a different behavior; the wall-normal velocity and polymer force are anti-correlated almost everywhere, especially in the bulk of  the flow, while they are strongly correlated in the four corners, where the polymer in-plane stress is acting to enhance the turbulent fluctuations.

Finally, we analyze the mean first normal stress difference $\mathcal{N}_1=\sigma_{11}-\sigma_{22}$, being $\sigma_{11}$ and $\sigma_{22}$ the mean streamwise and wall-normal total stresses, \ie the sum of the solvent and polymer stresses. The first normal stress difference $\mathcal{N}_1$ averaged over the whole domain is positive whereas the second one $\mathcal{N}_2$ is slightly negative, with $\vert \mathcal{N}_1/\mathcal{N}_2 \vert \gg 1$. The spatial distribution of $\mathcal{N}_1$ is shown on the right panel of \figrefA{fig:cijCorr}; $\mathcal{N}_1$ is greater than zero in the whole domain and its peaks are located at the wall, around $0.5h$ from the side walls. Also, the first normal stress difference is close to zero at the centerline and in the four corners. Note that, as expected the described spatial distribution of $\mathcal{N}_1$ closely follows that of $C_{11}$ previously discussed in \figrefA{fig:cij}.

\subsection{Flow structures}
\begin{figure}
\centering
\includegraphics[width = 0.75\textwidth]{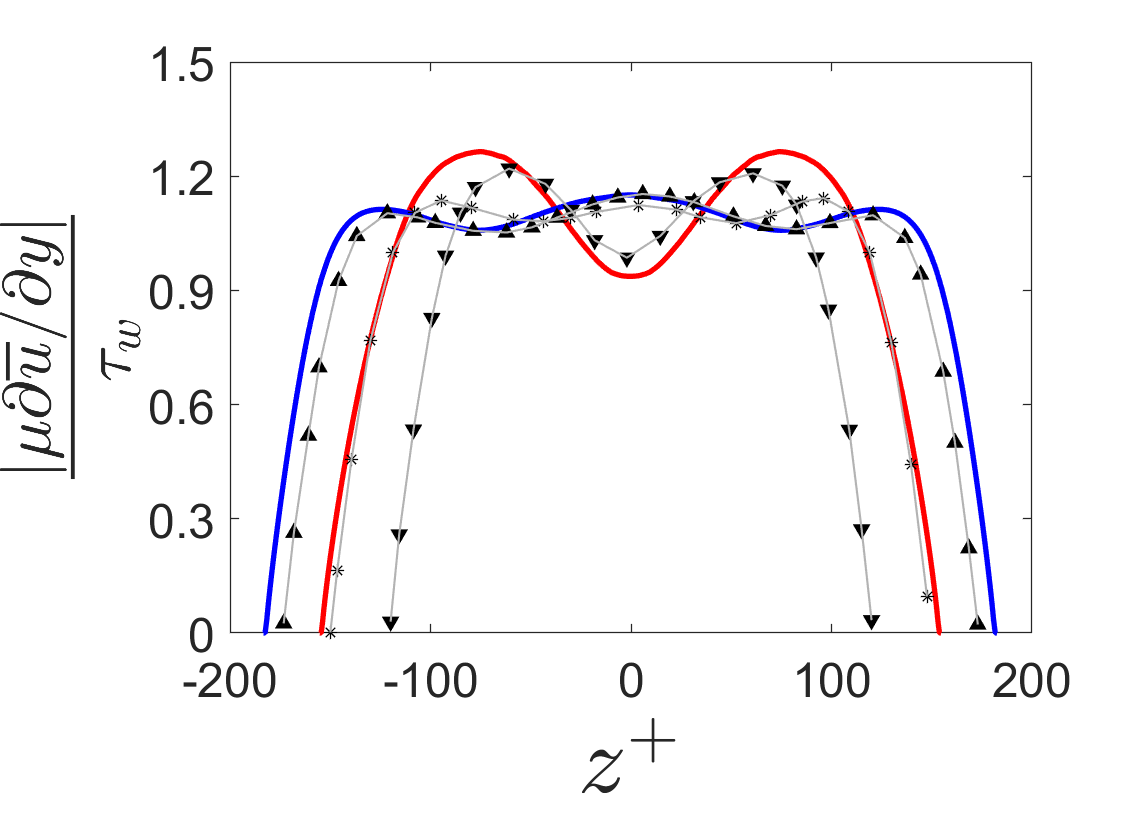}
\caption{Mean local wall stress along one edge in wall coordinates normalized by the average over the whole wall $\tau_w$. The blue and red lines are used to show our numerical results of the Newtonian and polymeric flows, respectively, while the grey lines with symbols are the data taken from the work of \citet{pinelli_uhlmann_sekimoto_kawahara_2010a} for three Newtonian flows with different Reynolds numbers: $Re_\tau=178$ ($\blacktriangle$), $150$ ($\star$) and $120$ ($\blacktriangledown$).}
\label{fig:tauwall}
\end{figure}
To further characterize the flow we study the mean local wall stress along one edge in wall coordinates normalized by the average over the whole wall, similarly to what was proposed by \citet{pinelli_uhlmann_sekimoto_kawahara_2010a}. These authors suggested that an upper bound for the number of wall velocity streaks over one of the four walls can be estimated from the dimension of the square expressed in wall units, since the average distance between streaks of different velocity sign is of the order of $50^+$ \citep{kim_moin_moser_1987a}. In our case, we can expect more streaks in the Newtonian case than in the viscoelastic flow with polymers since increasing streak spacing is a well-known effect of polymer drag reduction that has been observed both in experiment and numerical works \cite[\eg][]{de-angelis_casciola_piva_2002a}. Indeed, this argument is confirmed by \figrefA{fig:tauwall}, which shows that in the Newtonian case each edge of the duct hosts up to five streaks, with two high-velocity streaks close to the side walls and one in the center, and two low-velocity streaks in between, represented in the figure by the three maxima and two minima in the wall stress profile. In the drag-reduced case with polymers, the profile presents only two maxima and one minimum, indicating that two high-velocity streaks are close to the side walls and one low-velocity streak is located between them at the center of the edge; also, the two peaks are closer to the centerline than in the Newtonian case. In the figure, we also report with symbols  three curves pertaining to the results by \citet{pinelli_uhlmann_sekimoto_kawahara_2010a} for a Newtonian duct flow at $Re_\tau=178$ ($\blacktriangle$), $150$ ($\times$) and $120$ ($\blacktriangledown$): the first is used as reference, showing the good agreement between our Newtonian results and the one from the literature, while the other two are included to provide further insight on the polymeric flow. Indeed, the Newtonian case at $Re_\tau=150$ has approximately the same Reynolds number as our polymeric flow ($155$), but still shows a local maximum at the centerline, while the Newtonian case at $Re_\tau=120$ exhibits a pattern of minima and maxima similar to the one obtained by our simulation with polymers, even if with different (lower) Reynolds number. Thus, even at a similar friction Reynolds number, the shear stress profile does not resemble the one of a Newtonian flow due to the drag reducing effect of polymers.

\begin{figure}
\centering
\includegraphics[width = 0.74\textwidth]{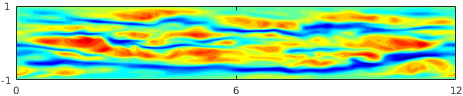}
\includegraphics[width = 0.21\textwidth]{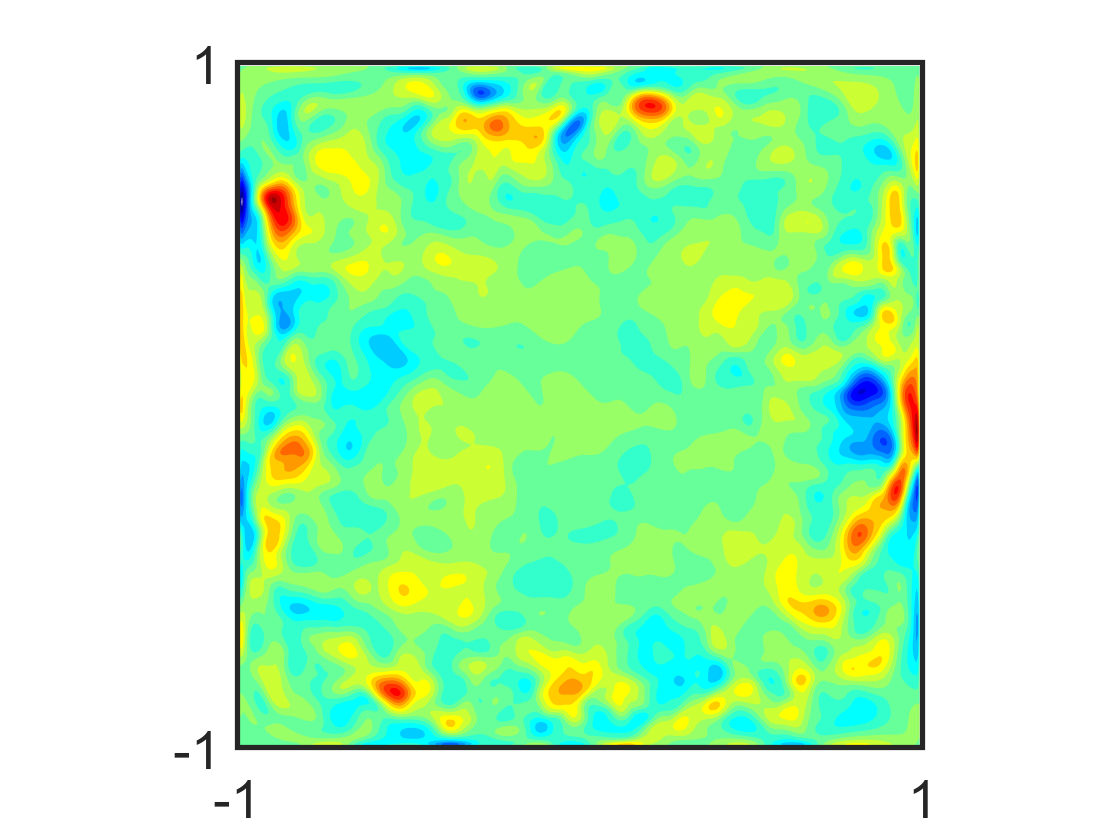}
\includegraphics[width = 0.74\textwidth]{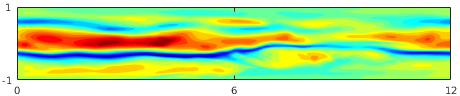}
\includegraphics[width = 0.21\textwidth]{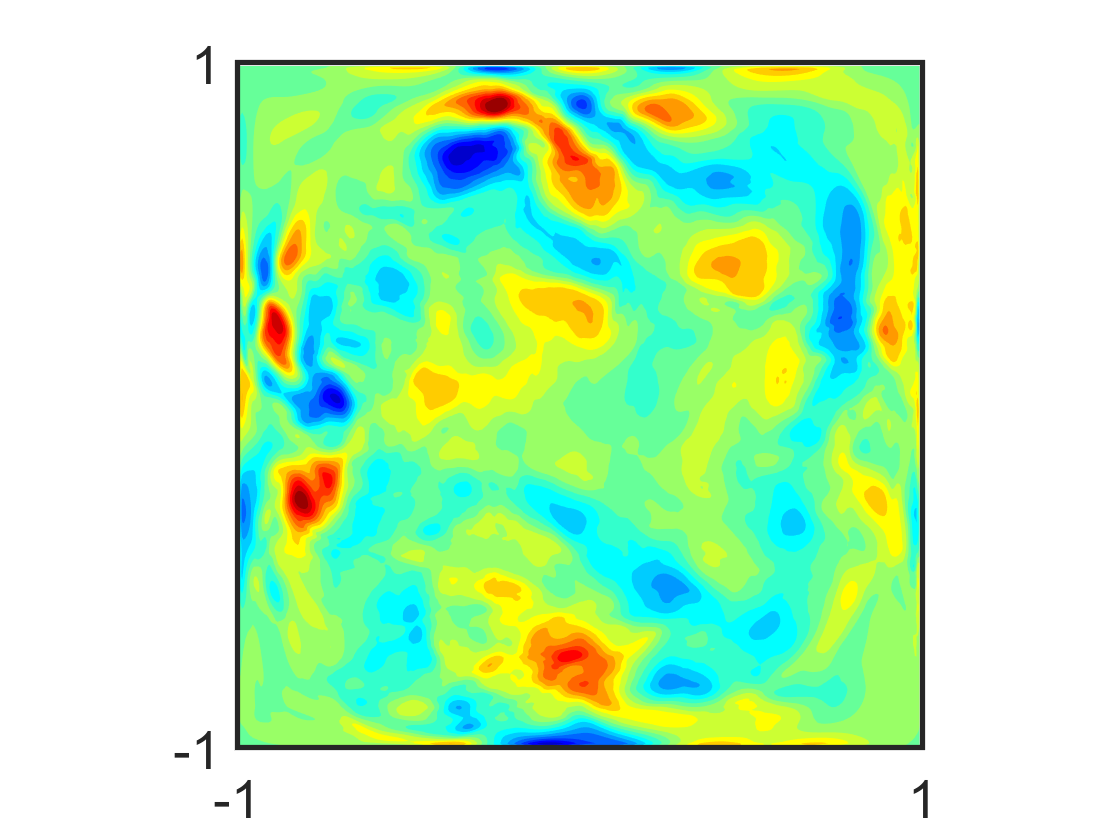}
\caption{Contour of the instantaneous (left) streamwise velocity fluctuation component $u'$ in the $x-z$ plane located at $\widetilde{y} \approx 18^+$ from the wall and (right) streamwise vorticity $\omega_x$ in the $x-y$ plane. The top and bottom rows show the Newtonian and polymer flows, respectively. The color scale ranges from $-0.5U_b$ (blue) to $0.5U_b$ (red).}
\label{fig:Vinst}
\end{figure}
The previous result is confirmed by \figref[left]{fig:Vinst}, which reports contours of the instantaneous streamwise velocity fluctuation $u'$ in a $x-z$ section located at a distance $\widetilde{y} \approx 18^+$ from one wall. The picture clearly shows the instantaneous footprints of the near-wall low- and high-speed streaks. Even from an instantaneous picture of the flow, the difference between the Newtonian and polymeric flow appears evident; indeed, the former shows two long low-speed streaks  and more than two high-speed streaks, which are shorter and wider than the low-speed ones, whereas the viscoelastic flow presents just one low-speed streak close to the centerline, bounded by two high-speed streaks. Further, both the low- and high speed streaks are stronger and correlated over longer distances in the streamwise direction in the presence of polymers \citep{warholic_massah_hanratty_1999a, escudier_nickson_poole_2009a}. \citet{pinelli_uhlmann_sekimoto_kawahara_2010a} showed that the mean streamwise vorticity (secondary flow) strongly depends upon the statistically preferred location of the quasi-streamwise vortices associated with the pair of fast/slow streaks closest to the corner. As just shown in \figrefA{fig:tauwall} and \figrefA{fig:Vinst}, the inclusion of the polymers in the flow strongly modifies these structures, by enhancing the streamwise coherence of the flow and by increasing the size of the near wall structures (see \figref[right]{fig:Vinst}), thus moving the quasi-streamwise vortices away from the bottom and side walls: therefore, the secondary motion is characterized by large structures close to the center of the duct, due to the constraint imposed by the size of the domain.

\subsection{Vorticity budget}
\begin{figure}
\centering
\includegraphics[width = 0.75\textwidth]{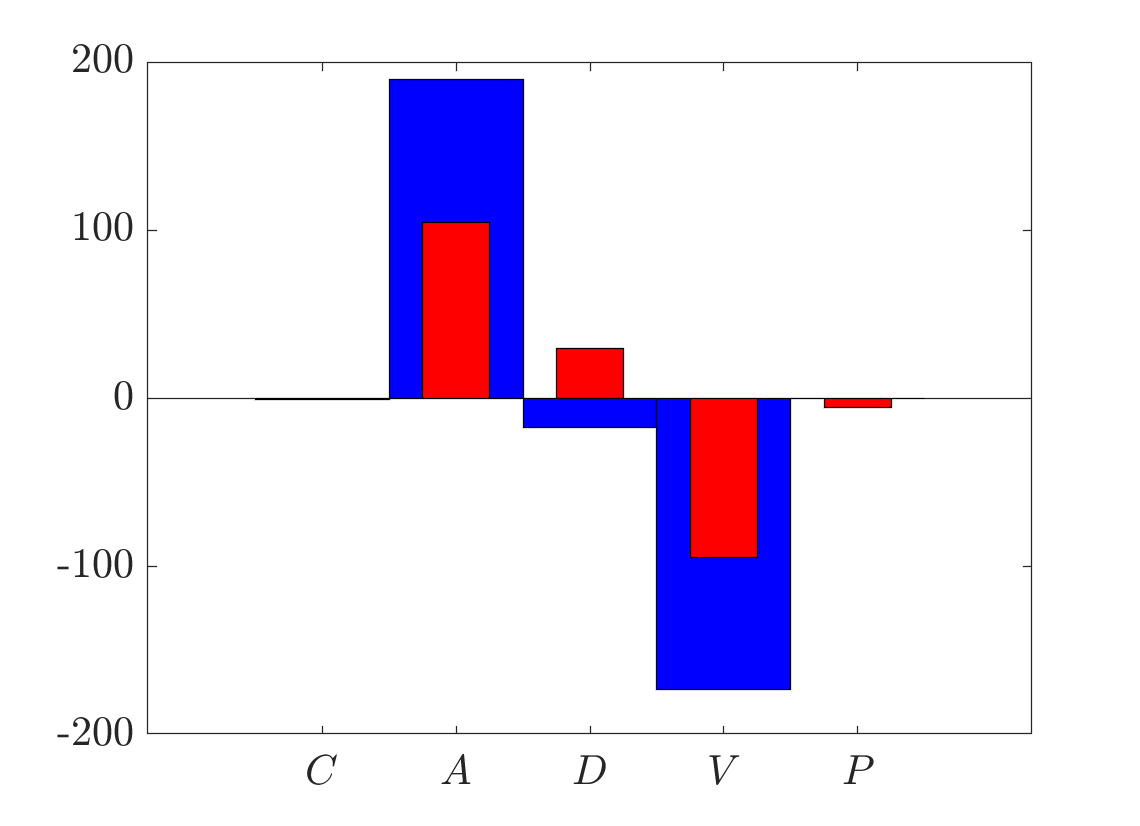}
\caption{Contributions to the budget of the mean streamwise vorticity $\overline{\omega}_x$: convection $\mathcal{C}$, production due to anisotropy of the Reynolds stress $\mathcal{A}$, production/dissipation source term $\mathcal{D}$, viscous dissipation $\mathcal{V}$, polymeric contribution $\mathcal{P}$. Blue and red colors are used for the Newtonian and viscoelastic flows, respectively.}
\label{fig:vorticityBalance}
\end{figure}

\citet{gavrilakis_1992a} explains the origins of the mean secondary flow using the equation for the mean streamwise vorticity $\overline{\omega}_x$, that we here modify for a fully developed duct flow with polymers and whose derivation has been reported in the Appendix. The equation for the streamwise vorticity reads
\begin{multline} \label{eq:vorticityBalance}
\frac{\partial \overline{\omega}_x}{\partial t} = \underbrace{\overline{v} \frac{\partial \overline{\omega}_x}{\partial y} +\overline{w} \frac{\partial \overline{\omega}_x}{\partial z}}_{\mathcal{C}} + \underbrace{\frac{\partial^2}{\partial y \partial z} \left( \overline{w' w'} -\overline{v' v'} \right)}_{\mathcal{A}} + \underbrace{\left( \frac{\partial ^2}{\partial y^2} - \frac{\partial ^2}{\partial z^2} \right) \overline{v'w'}}_{\mathcal{D}} + \\ ~ \\
\underbrace{-\frac{\beta}{Re} \left( \frac{\partial ^2}{\partial y^2} + \frac{\partial ^2}{\partial z^2} \right) \overline{\omega}_x}_{\mathcal{V}} ~\underbrace{-\frac{1 - \beta}{Re} \left( \frac{\partial ^2 \overline{\tau}_{32}}{\partial y^2} + \frac{\partial ^2 \overline{\tau}_{33}}{\partial z \partial y} - \frac{\partial ^2 \overline{\tau}_{22}}{\partial y \partial z} - \frac{\partial ^2 \overline{\tau}_{23}}{\partial z^2} \right) }_{\mathcal{P}}=0.
\end{multline}

The first two terms in the equation (denoted as $\mathcal{C}$) are the convection of the mean vorticity by the secondary flow itself, while the last two terms are a sink of streamwise vorticity due to the viscosity ($\mathcal{V}$) and the polymer contribution ($\mathcal{P}$), which we found to be a sink as well. The two remaining terms involve the Reynolds stresses: the first is a production term associated with the anisotropy of the in-plane normal stresses ($\mathcal{A}$), while the second ($\mathcal{D}$) can be both a production and dissipation term due to the cross Reynolds stress component $\overline{v'w'}$. \figrefAC{fig:vorticityBalance} shows the contribution of the five terms highlighted in \equref{eq:vorticityBalance}, \ie $\mathcal{C}$, $\mathcal{A}$, $\mathcal{D}$, $\mathcal{V}$, $\mathcal{P}$, in one of the $8$ sectors and averaged among all the sectors. Blue and red colors are used for the Newtonian and polymeric flows, respectively. The only positive contribution in the Newtonian case is provided by the anisotropic terms, while all the others are negative. With polymers, the polymeric contribution $\mathcal{P}$  is found to be a sink of vorticity, the anisotropic and viscous terms decreases in magnitude compared to the Newtonian flow, while $\mathcal{D}$ changes sign and increases in magnitude, changing from dissipation for the Newtonian flow to production for the flow with polymers. Finally, the convection term $\mathcal{C}$ is null for both cases. The budget analysis suggests that, although the polymeric stress is small in amplitude, its presence modifies the flow itself, by reducing the anisotropy of the in-plane velocity fluctuations.
 
\begin{figure}
\centering
\includegraphics[width = 0.23\textwidth]{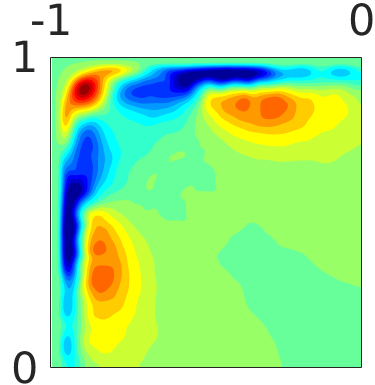}
\includegraphics[width = 0.23\textwidth]{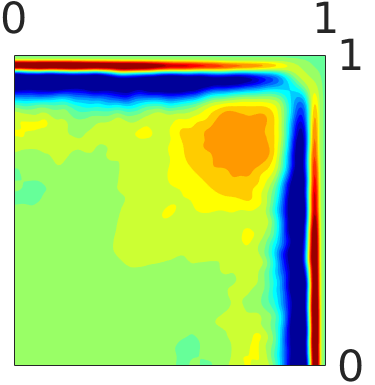} \hspace{0.25cm}
\includegraphics[width = 0.23\textwidth]{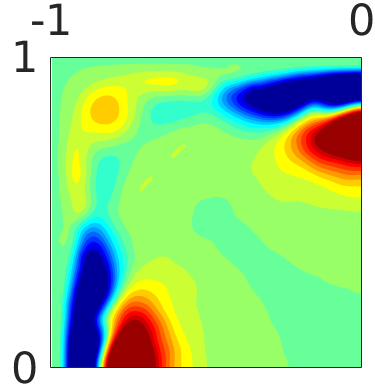}
\includegraphics[width = 0.23\textwidth]{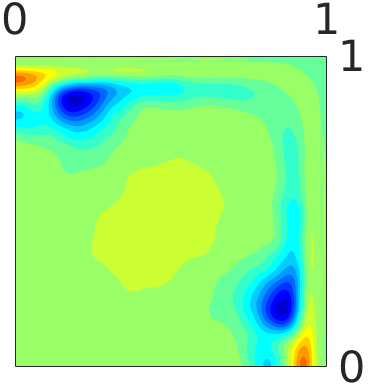}
\includegraphics[width = 0.23\textwidth]{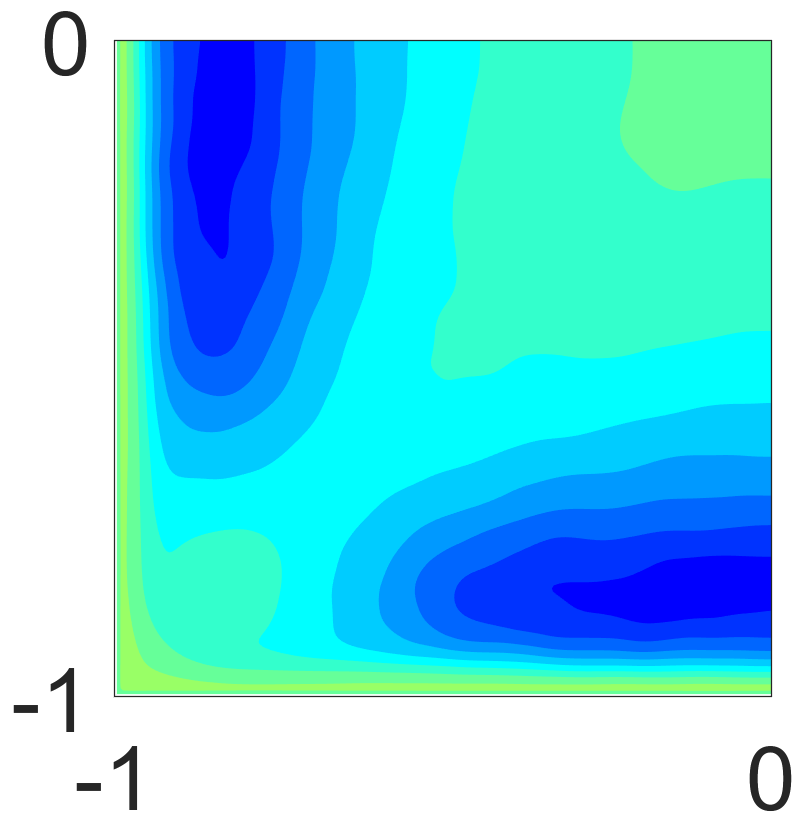}
\includegraphics[width = 0.23\textwidth]{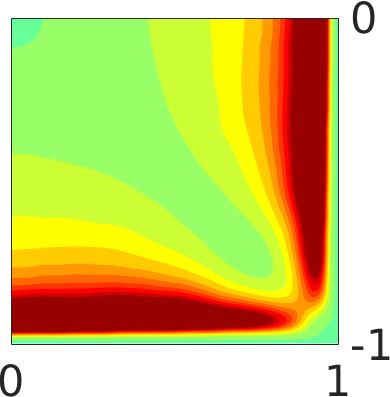} \hspace{0.25cm}
\includegraphics[width = 0.23\textwidth]{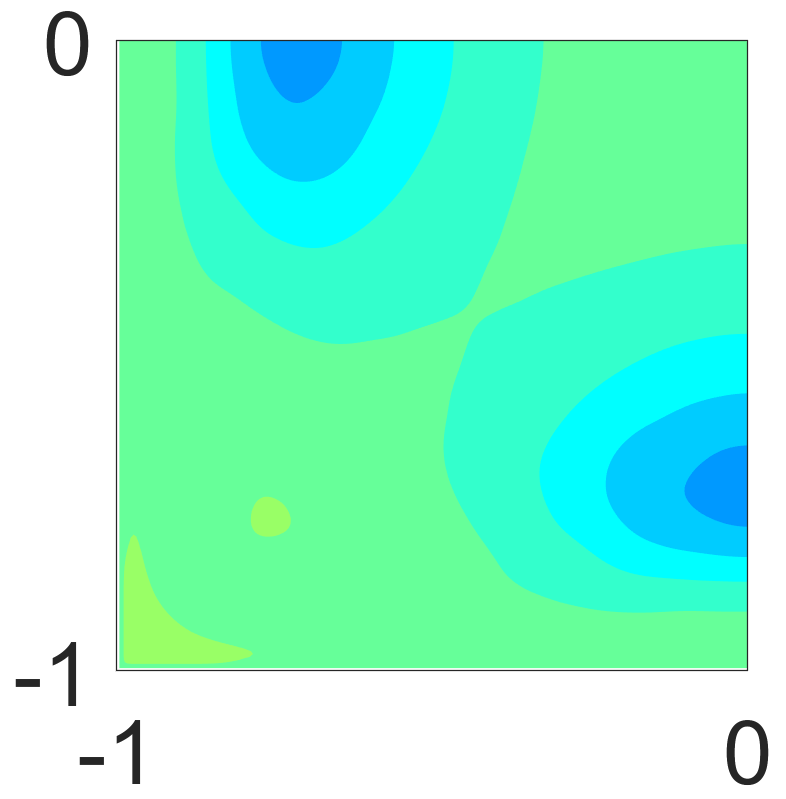}
\includegraphics[width = 0.23\textwidth]{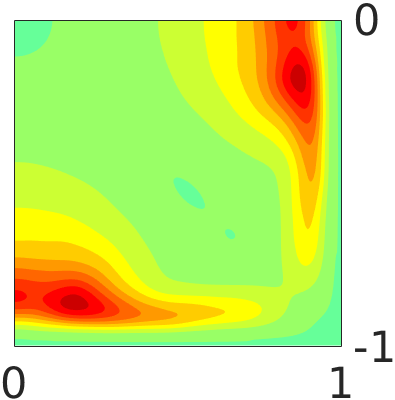} \\ \vspace{0.75cm}
\includegraphics[width = 0.23\textwidth]{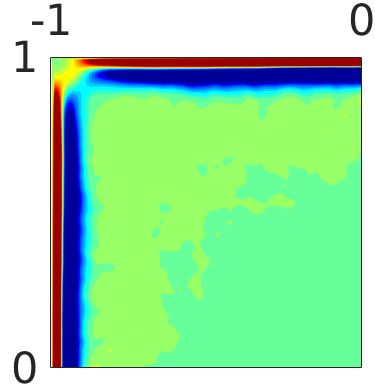}
\includegraphics[width = 0.23\textwidth]{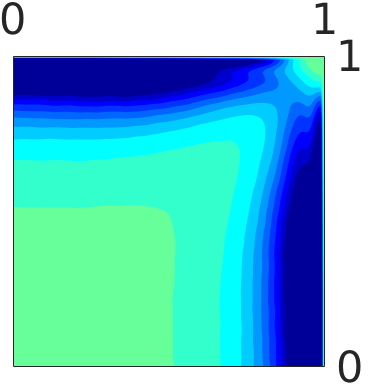} \hspace{0.25cm}
\includegraphics[width = 0.23\textwidth]{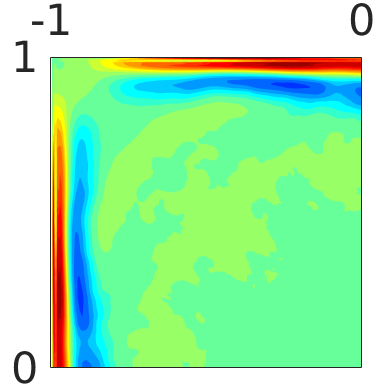}
\includegraphics[width = 0.23\textwidth]{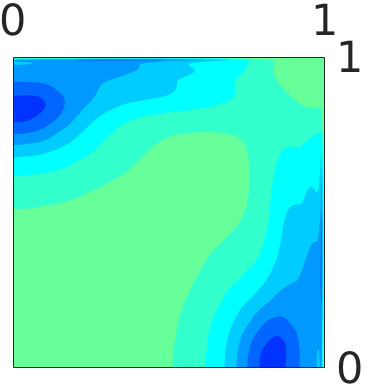}
\includegraphics[width = 0.23\textwidth]{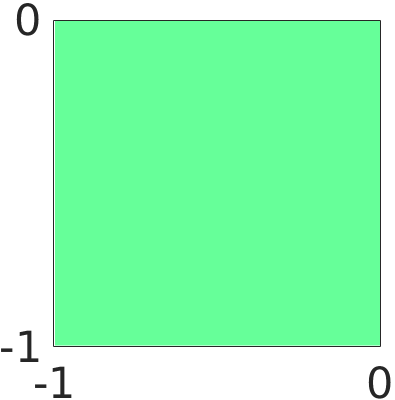}
\includegraphics[width = 0.23\textwidth]{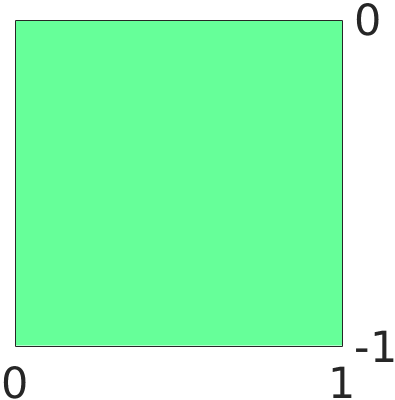} \hspace{0.25cm}
\includegraphics[width = 0.23\textwidth]{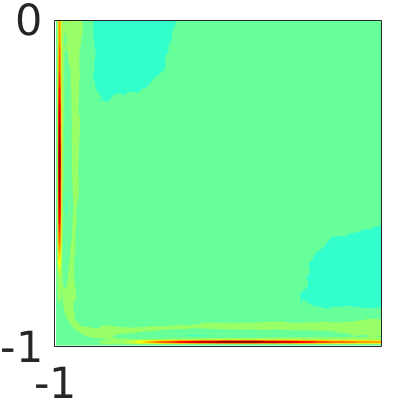}
\includegraphics[width = 0.23\textwidth]{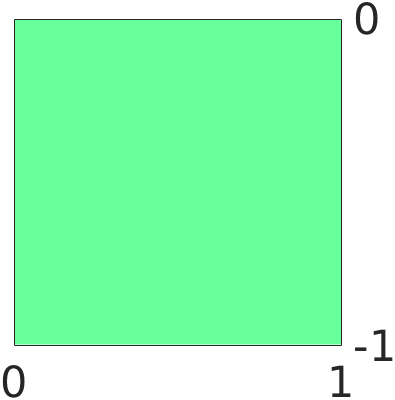}
\caption{Contributions to the streamwise Reynolds-stress tensor $\overline{u'u'}$ equation for the Newtonian (left panels) and viscoelastic case with polymers (right panels). On each side, the contours represent the terms $\mathcal{A}_{ij}$, $\mathcal{Q}_{ij}$, $\mathcal{R}_{ij}$, $\mathcal{P}_{ij}$, $\mathcal{D}_{ij}$, $\mathcal{\epsilon}_{ij}$, $\mathcal{W}_{ij}$, and the total balance (see \equref{eq:energyBalance}) in order from left to right and top to bottom. The ranges of the color scales are $\pm 0.002$, $\pm 0.008$, $\pm 0.005$, $\pm0.025$, $\pm 0.01$, $\pm 0.02$, $\pm 0.003$ and $\pm 0.001$, respectively, with colors going from blue (negative values) to red (positive value) and with the zero colored in green.}
\label{fig:energyBalance}
\end{figure}

\subsection{Energy budget}
To better understand the effect of polymers on the streamwise velocity fluctuations $\overline{u'u'}$, which are related to steaks, we consider the perturbation energy budget. The transport equation for the Reynolds stress in a viscoelastic flow is given by 
\begin{multline} \label{eq:energyBalance}
\frac{\partial \overline{ u'_i u'_j}}{\partial t}= \underbrace{- \overline{u}_k \frac{\partial \overline{ u'_i u'_j}}{\partial x_k}}_{\mathcal{A}_{ij}} \underbrace{- \overline{u}_k \frac{\partial \overline{ u'_i u'_j u'_k}}{\partial x_k}}_{\mathcal{Q}_{ij}} \underbrace{- (\overline{u'_j \frac{\partial p'}{\partial x_i}}+\overline{u'_i \frac{\partial p'}{\partial x_j}} )}_{\mathcal{R}_{ij}}  \underbrace{- (\overline{u'_j u'_k }\frac{\partial \overline{u}_i}{\partial x_k}+\overline{u'_i u'_k }\frac{\partial \overline{u}_j}{\partial x_k})}_{\mathcal{P}_{ij}} + \\
+ \underbrace{\frac{\beta}{Re}\frac{\partial ^2 \overline{ u'_i u'_j} }{\partial x^2_k}}_{\mathcal{D}_{ij}} \underbrace{-\frac{2\beta}{Re} \overline{ \frac{\partial  u'_i  }{\partial x_k} \frac{ u'_j  }{\partial x_k}}}_{\mathcal{\epsilon}_{ij}}+\underbrace{\frac{1-\beta}{Re}( \overline{u'_j\frac{\partial \tau'_{ik}}{\partial x_k}}+\overline{u'_i\frac{\partial \tau'_{jk}}{\partial x_k}})}_{\mathcal{W}_{ij}}=0
\end{multline}
where $\mathcal{A}_{ij}$ is the advection by the mean flow, $\mathcal{Q}_{ij}$ the transport by the velocity fluctuations, $\mathcal{R}_{ij}$ the pressure term, $\mathcal{P}_{ij}$ the production against the mean shear, $\mathcal{D}_{ij}$ the viscous diffusion, $\mathcal{\epsilon}_{ij}$ the dissipation, and $\mathcal{W}_{ij}$ the polymer work. The evolution equation for the perturbation energy is obtained by setting $i = j$, and in particular we consider here the streamwise component $i = j = 1$. \figrefAC{fig:energyBalance} shows the contours of these seven contributions, and their sum, for the Newtonian (left panel) and polymeric flow (right panel). First, we note that, as expected, the sum of all the contribution is in both cases almost zero (up to $10^{-3}$), since the mean Reynolds stress $\overline{u'u'}$ has reached steady state.

We first discuss  the Newtonian flow, reported on the left panels; the advection $\mathcal{A}_{ij}$ and transport $\mathcal{Q}_{ij}$ terms display strong gradients close to the walls. In particular, the distribution of $\mathcal{A}_{ij}$ reveals that the maximum advection occurs in regions close to the maximum in-plane mean vorticity, \ie where the secondary flow is maximum (see \figrefA{fig:mean}). On the other hand, the turbulent transport, $\mathcal{Q}_{ij}$, is observed to be approximately uniform along the edges; in particular, it is zero at the wall, reaches a positive peak value in the near wall region, followed by a minimum and finally goes back to zero away from the wall, thus indicating that the largest gradients of the high-order statistics are located mainly in the near wall region, all along the edges. The pressure and production terms, $\mathcal{R}_{ij}$ and $\mathcal{P}_{ij}$, exhibit opposite behaviour: the former is always negative and the latter always positive, reaching a minimum/maximum and then approaching zero again in the bulk of the duct. The viscous diffusion $\mathcal{D}_{ij}$ displays a strong peak at the wall and then decreases to zero, while the dissipation term $\mathcal{\epsilon}_{ij}$ has a minimum at the wall, and then increases to its maximum value in the center of the duct. The figure shows that all the contributions have an almost uniform distribution along the edges, with only a weak dependency on the distance from the corners, except for the advection $\mathcal{A}_{ij}$ and transport $\mathcal{Q}_{ij}$ terms.

The behavior of the different terms in the kinetic energy budget is different for the viscoelastic flow, as shown in the right panel of \figrefA{fig:energyBalance}. In this case, all the terms are not uniform along the edge and strongly depend on the distance from the side walls; this is related to the reduced friction Reynolds number of the flow and the consequent increase in size of the near wall structures previously discussed. In particular, the positive and negative peaks of the advection $\mathcal{A}_{ij}$ term are found close to the centerline far from the corners, similarly to the transport term $\mathcal{Q}_{ij}$ which exhibits strong negative and positive peaks in the same region. Also, the pressure $\mathcal{R}_{ij}$, production $\mathcal{P}_{ij}$ and dissipation $\mathcal{\epsilon}_{ij}$ terms exhibit similar trend, with maxima and minima located away from the corner. Finally, the polymer contribution ($\mathcal{W}_{ij}$) is not null, but its amplitude is small compared to the other terms, and its distribution across the whole duct is almost uniform. These results, clearly indicate that the polymers strongly modify the flow; in particular, their presence induce a less uniform flow in the duct, with most of the streamwise flow fluctuations displaced towards the centerline and away from the corner, as we have already noticed in \figrefA{fig:cijCorr}. The location where on average the low- and high-speed streaks are present, as previously shown in \figrefA{fig:tauwall} and \figrefA{fig:Vinst}, coincides with the location of major activity of the flow where strong peaks of advection, transport, production and dissipation are present.

\subsection{Elasticity effect}
\begin{table}
\centering
\setlength{\tabcolsep}{5pt}
\begin{tabular}{lrrrrr}
Case							& $Re$			& $Wi$		& $\beta$		& $L^2$	& $DR\%$		\\
P								& $2800$		& $1.5$		& $0.90$		& $3600$	& $29\%$		\\
P-$Wi\uparrow$		& $2800$		& $3.0$	& $0.90$		& $3600$	& $43\%$		\\
P-$L\downarrow$		& $2800$		& $1.5$		& $0.90$		& $~900$	& $27\%$		\\
P-$L\uparrow$			& $2800$		& $1.5$		& $0.90$		& $5625$	& $29\%$		\\
P-$\beta\uparrow$	& $2800$		& $1.5$		& $0.95$		& $3600$	& $25\%$		\\
\end{tabular}
\caption{Summary of all the DNSs performed at a fixed bulk Reynolds number equal to $Re=2800$ and with different viscoelastic fluids.}
\label{tab:cases}
\end{table}
We now assess the effect of the parameters defining the viscoelastic behaviour of the fluid. \tabrefC{tab:cases} reports all the different simulations that we have performed and the drag reduction $DR$ obtained in each case. In particular, starting from the reference polymeric case discussed before (P), we have varied independently the Weissenberg number $Wi$ (P-$Wi\uparrow$), the dumbbell extensibility $L$ (P-$L\downarrow$ and P-$L\uparrow$) and the viscosity ratio $\beta$ (P-$\beta\uparrow$). From the table we can observe that starting from the drag reduction reported for the reference polymeric case ($29\%$), a relevant reduction or increase of $L$ brings only small variations in the overall drag reduction ($27\%$ and $29\%$), while even smaller variation of $\beta$ are sufficient to alter the solution, as demonstrated by a drag reduction of $25\%$ for the case with higher viscosity ratio P-$\beta\uparrow$. However, the parameter that has the greatest effect on the solution appears to be the Weissenberg number $Wi$. Indeed, when $Wi=3$ (P-$Wi\uparrow$) the drag reduction increases to $43\%$. We will now analyse in more detail this polymeric case with $Wi=3$ and compare the flow with the one discussed above, \ie the Newtonian and the polymeric with $Wi=1.5$.

\begin{figure}
\centering
\includegraphics[height = 0.35\textwidth]{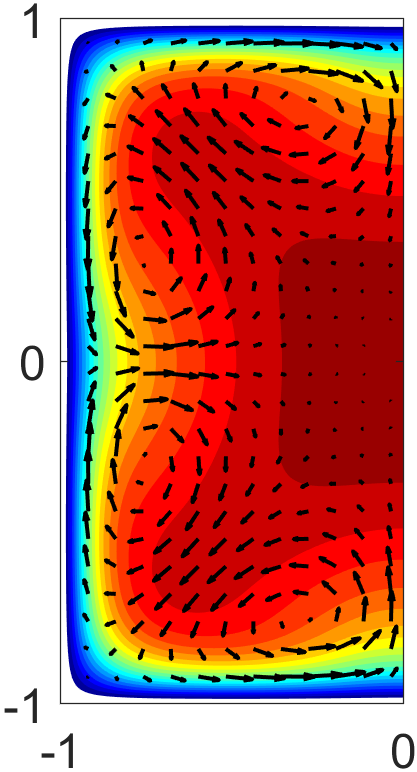}
\includegraphics[height = 0.35\textwidth]{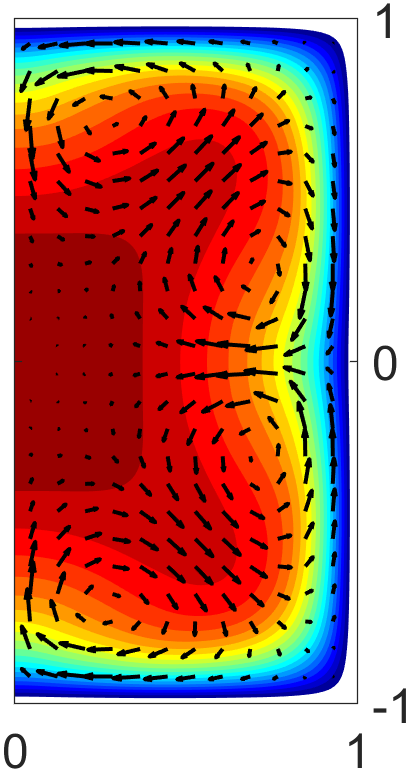}
\includegraphics[height = 0.35\textwidth]{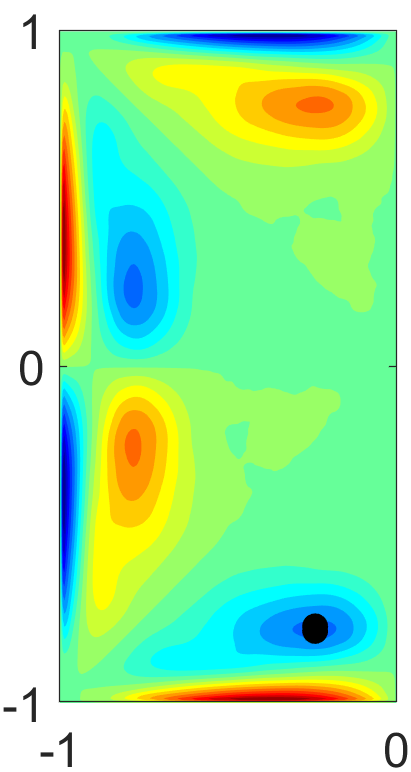}
\includegraphics[height = 0.35\textwidth]{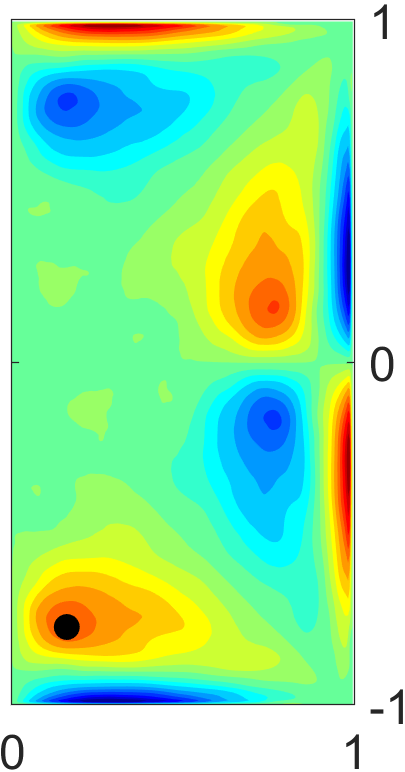}
\caption{Contour of the mean streamwise component of (left) velocity $\overline{u}$ and (right) vorticity $\overline{\omega}_x$. In each figure, the left and right half of the duct are used for the polymeric cases with $Wi=1.5$ and $Wi=3$, respectively. The color scale ranges from $0$ (blue) to $1.4U_b$ (red) in the left panel and from $-0.5U_b/h$ (blue) to $0.5U_b/h$ (red) in the right one. The two left panels also report the in-plane mean velocity with arrows.}
\label{fig:meanE}
\end{figure}
\figrefAC{fig:meanE} shows the contour of the mean streamwise components of the velocity $\overline{u}$ and vorticity $\overline{\omega}_x$ for the two different Weissenberg number considered. As $Wi$ increases, the maximum streamwise velocity is only slightly affected, while the maximum vorticity strongly decreases by $40\%$. However, the integral of the vorticity keeps increasing, thus confirming a progressive increase of the in-plane motion with the Weissenberg number $Wi$. The location of maximum vorticity, marked with a dot in the figure, moves further away from the walls, displacing towards the center; however, this displacement is smaller than what was previously observed when comparing the Newtonian case with the polymeric one with $Wi=1.5$ (see the right panel in \figrefA{fig:mean}), thus indicating a nonlinear trend with $Wi$.

\begin{figure}
\centering
\includegraphics[width = 0.325\textwidth]{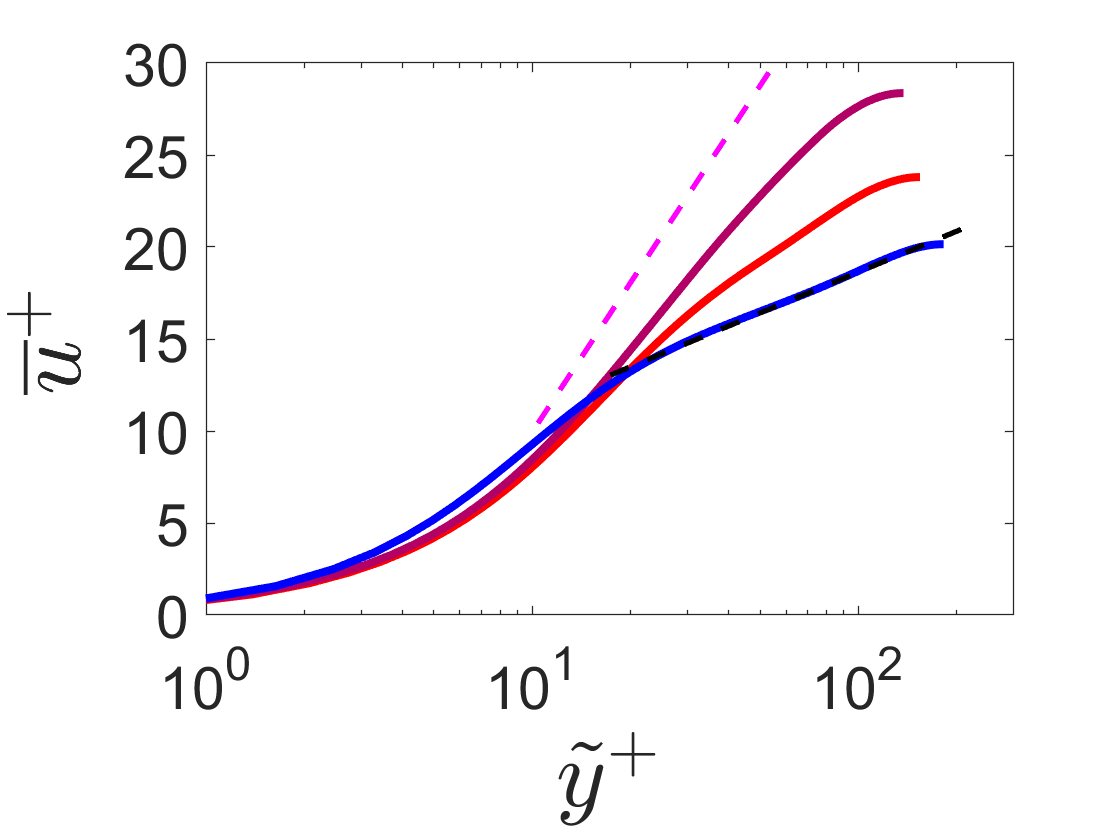}
\includegraphics[width = 0.325\textwidth]{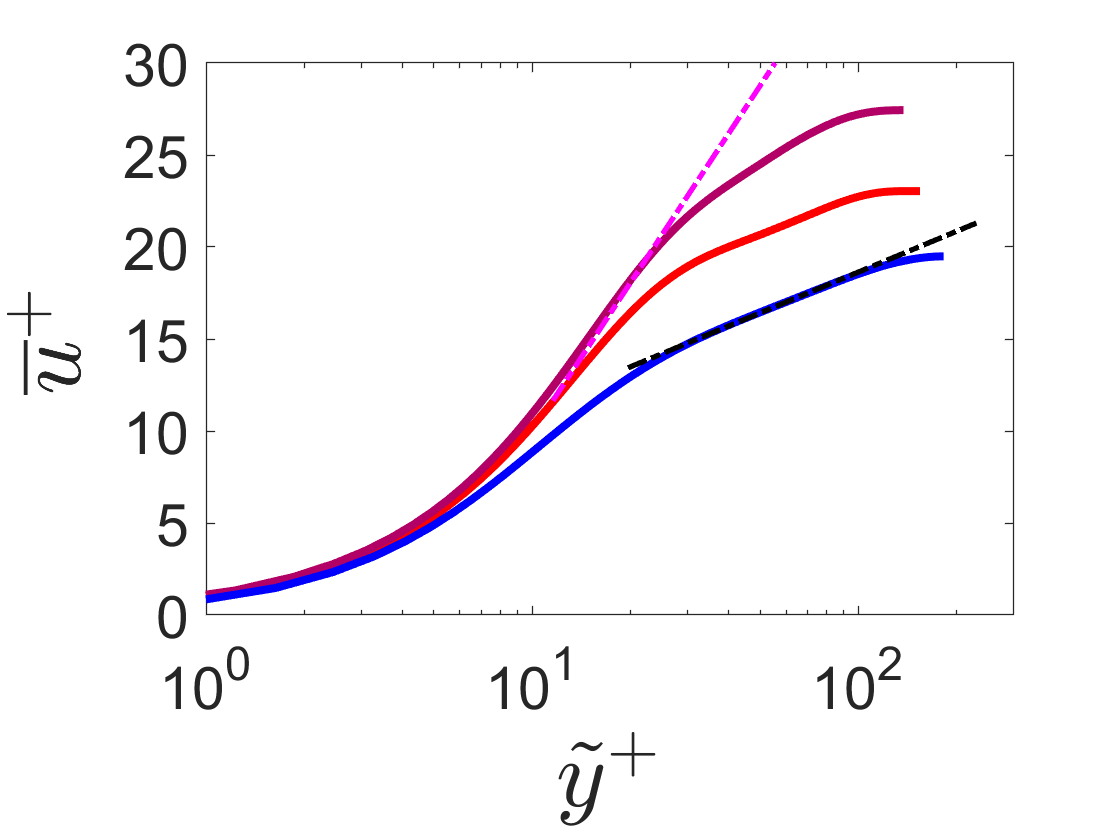}
\includegraphics[width = 0.325\textwidth]{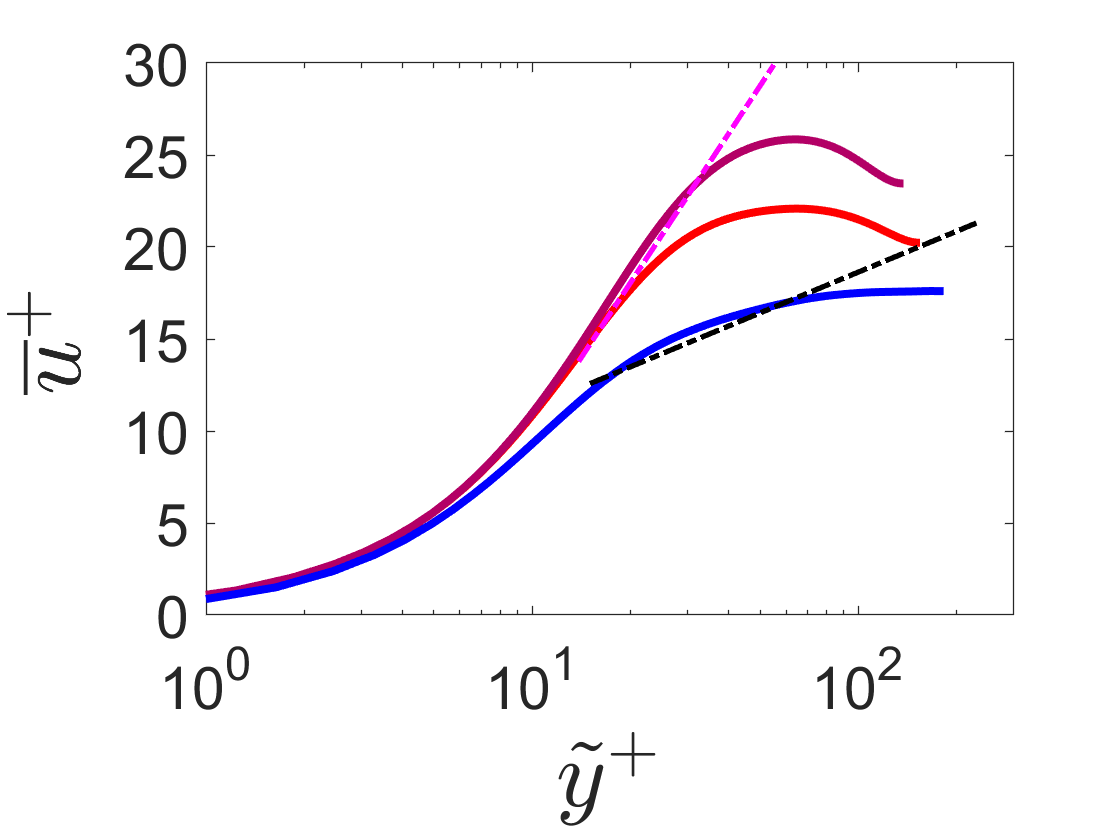}
\caption{Mean streamwise velocity as function of the wall-normal distance in logarithmic scale and wall units. The blue, red and brown lines are used for the numerical results without and with polymers with $Wi=1.5$ and $3$, respectively. The magenta dash dotted line is the polymer Maximum Drag Reduction \citep{virk_1975a, lvov_pomyalov_procaccia_tiberkevich_2004a} while the black line indicates the Newtonian log law. The three panels correspond to different sections: (left) $z=0$, (middle) $z=0.3h$, and (right) $z=0.6h$.}
\label{fig:loglawE}
\end{figure}
Next, we study the mean streamwise velocity profiles in wall units reported in \figrefA{fig:loglawE}. Here, we compare the profiles of the Newtonian fluid and the two viscoelastic cases with different Weissenberg number $Wi$. In all the considered sections, we observe a progressive departure from the Newtonian velocity profile as $Wi$ increases, with the slope of the inertial range of scales growing with $Wi$; the curves for $Wi=3$ approach the Maximum Drag Reduction limit.

\begin{figure}
\centering
\includegraphics[width = 0.325\textwidth]{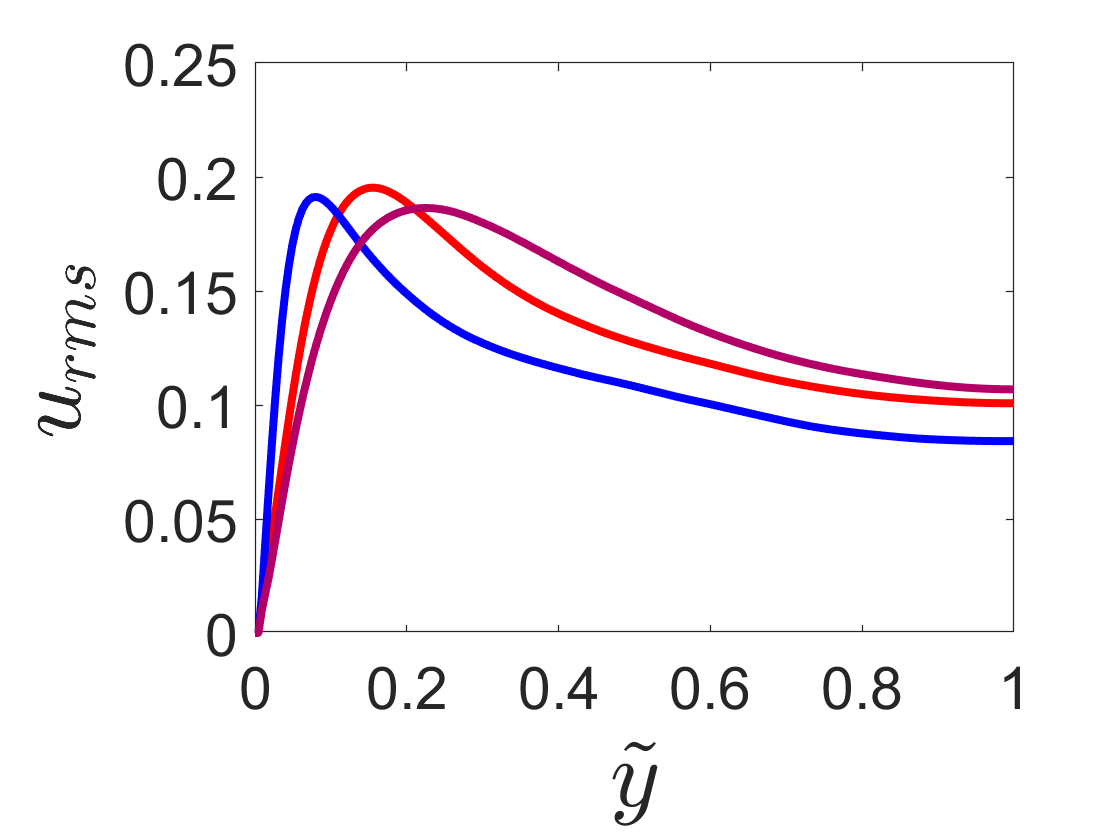}
\includegraphics[width = 0.325\textwidth]{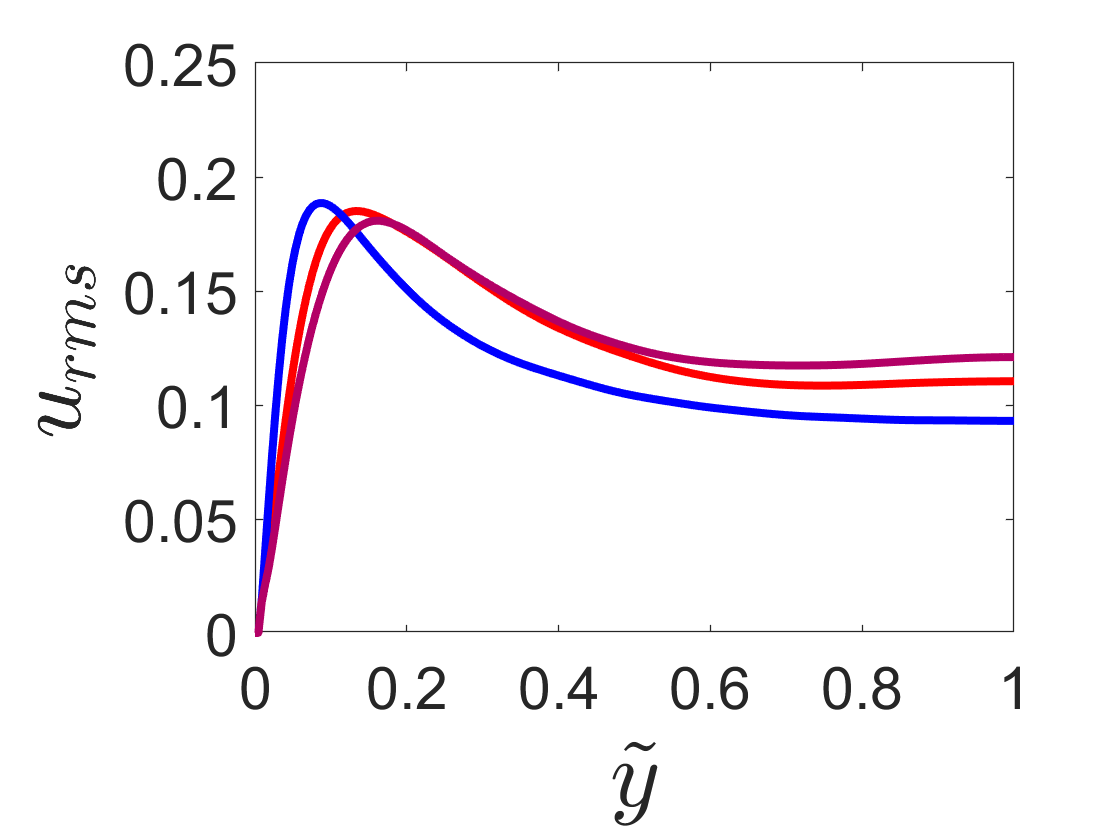}
\includegraphics[width = 0.325\textwidth]{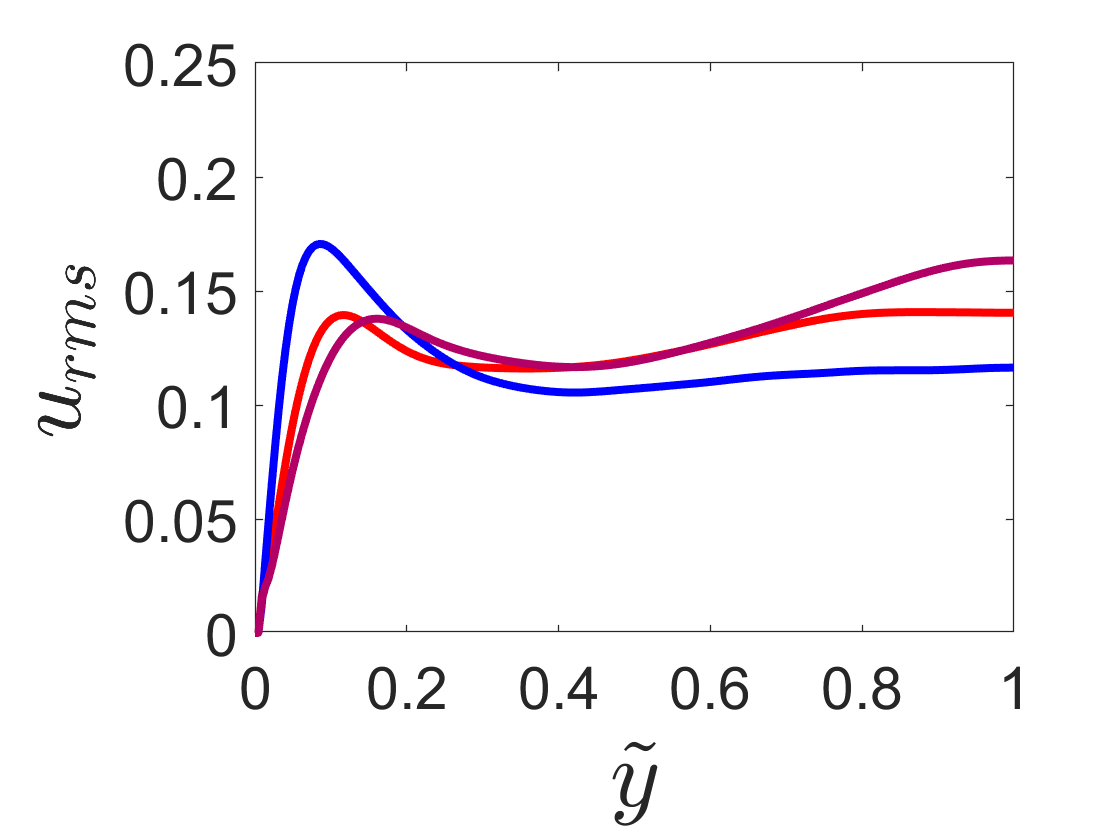}
\includegraphics[width = 0.325\textwidth]{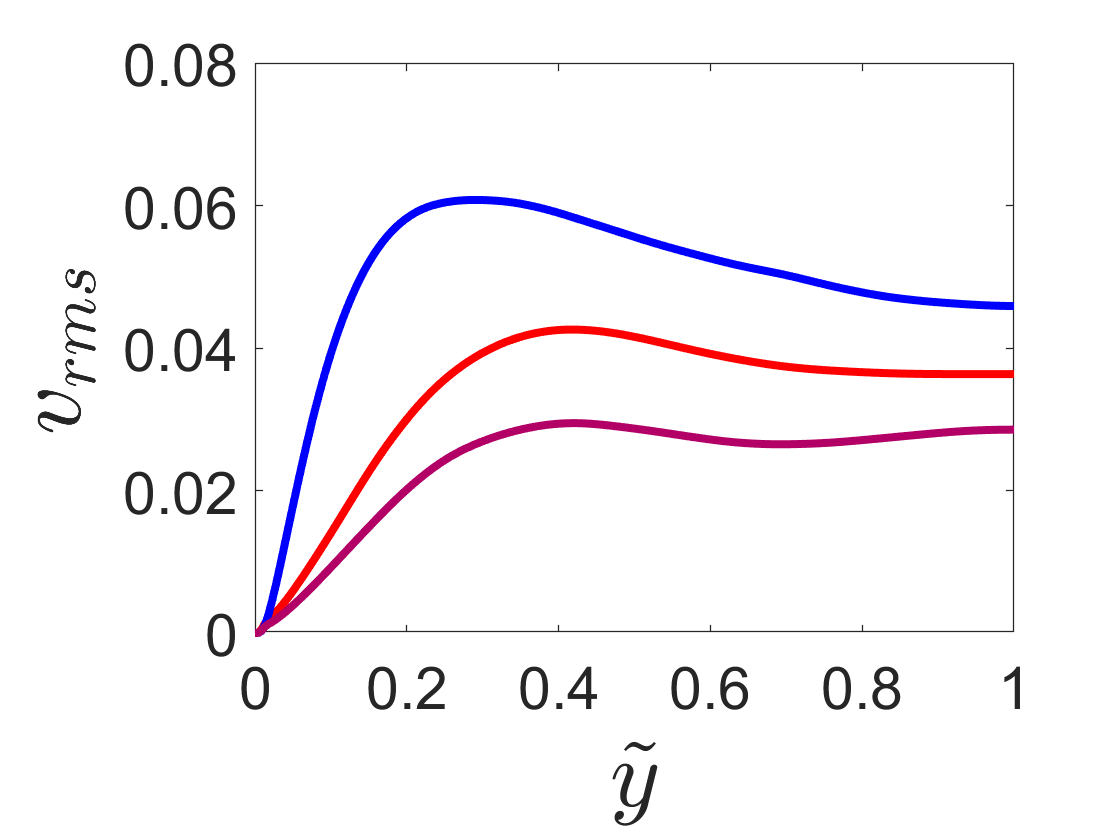}
\includegraphics[width = 0.325\textwidth]{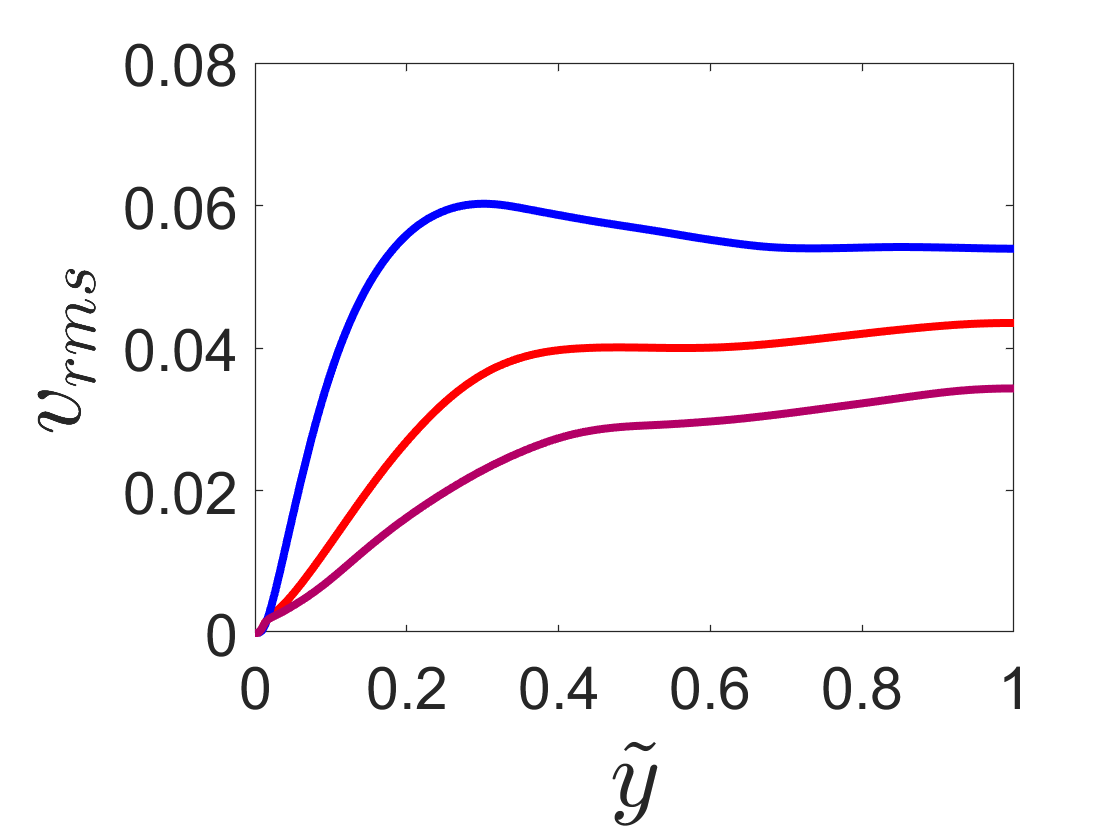}
\includegraphics[width = 0.325\textwidth]{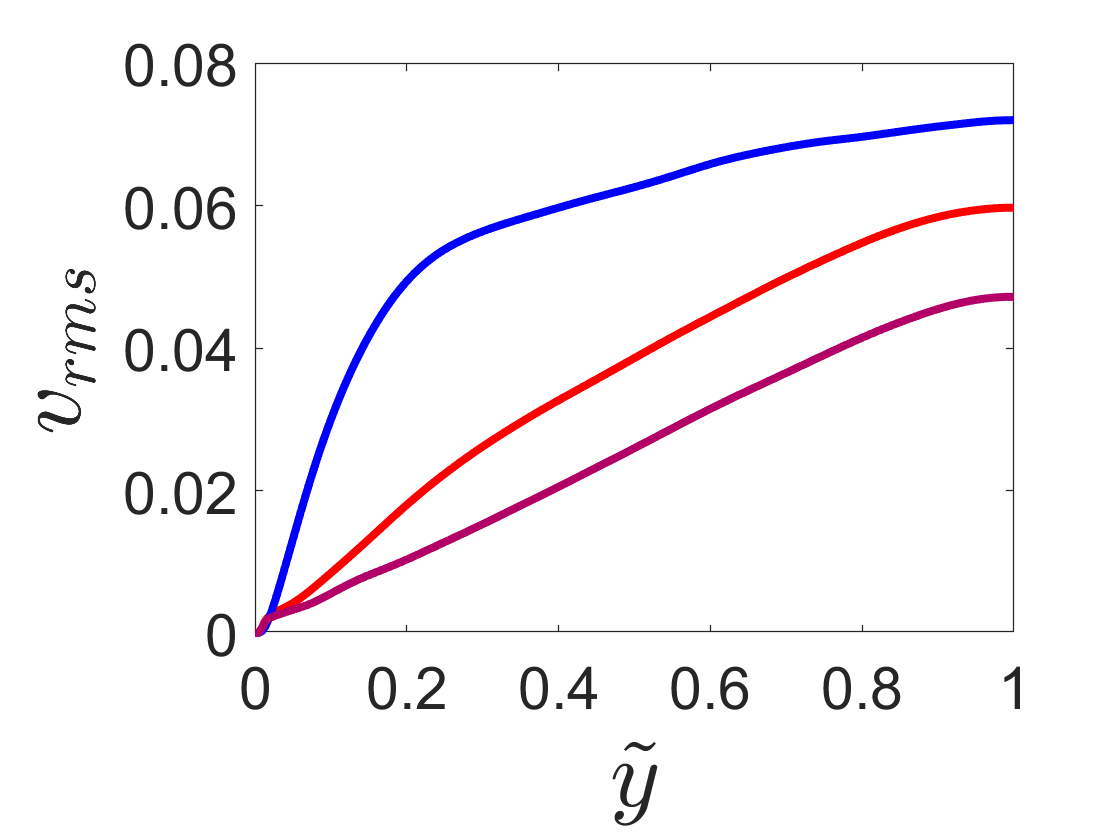}
\includegraphics[width = 0.325\textwidth]{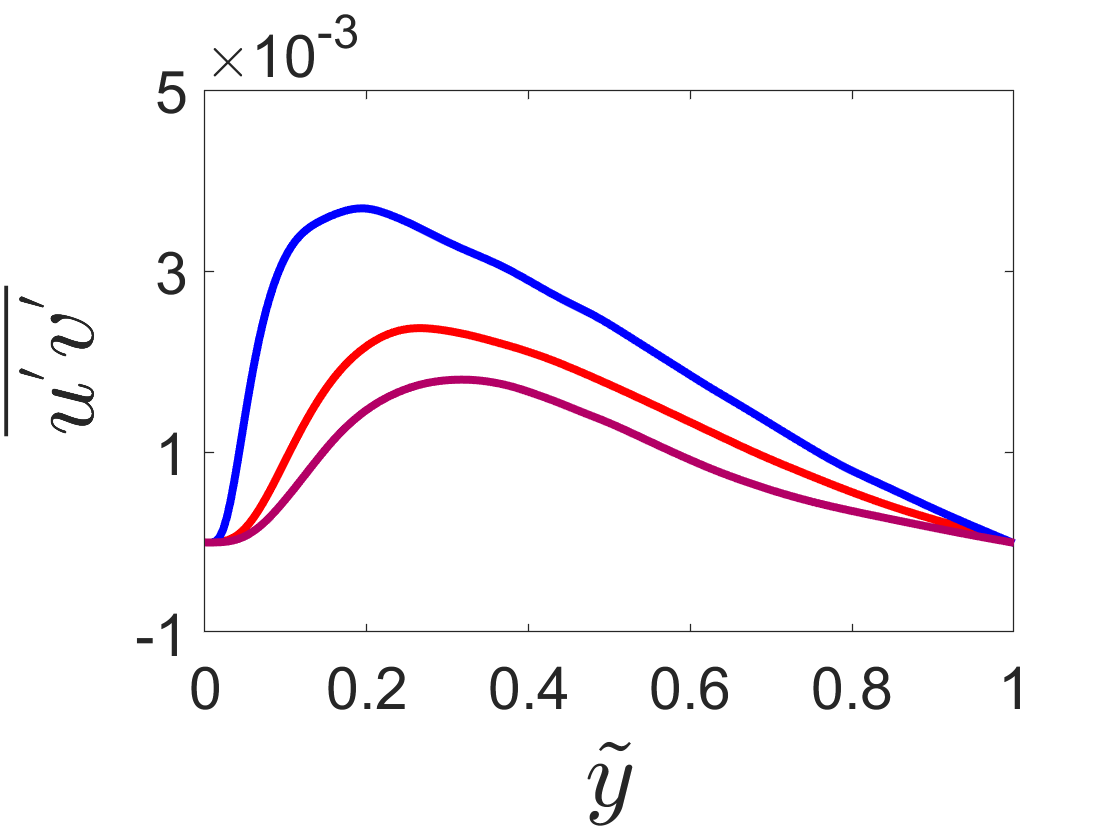}
\includegraphics[width = 0.325\textwidth]{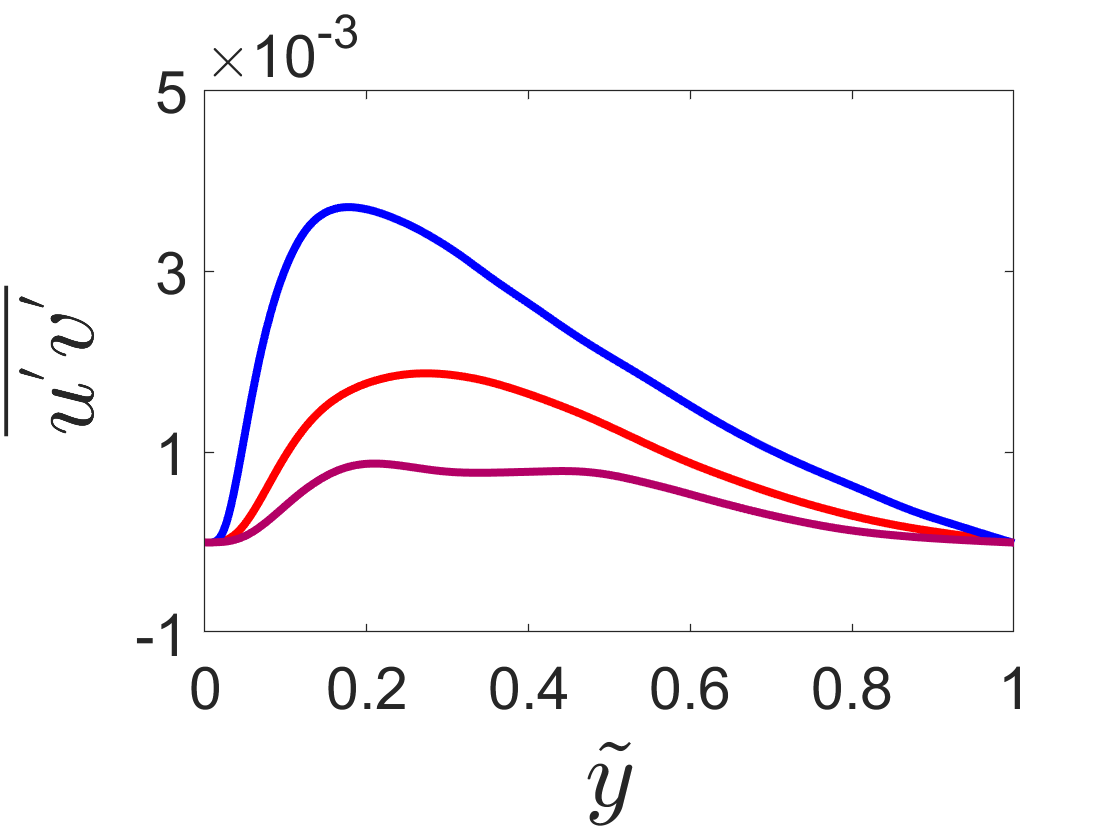}
\includegraphics[width = 0.325\textwidth]{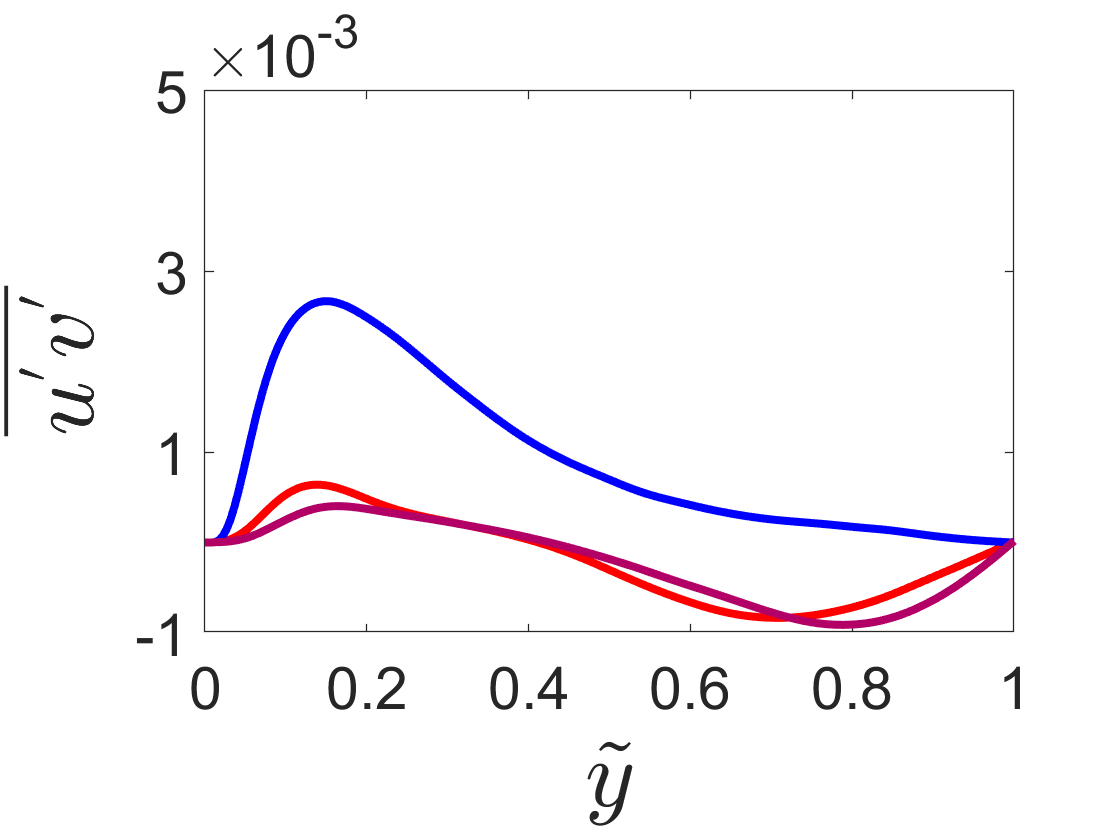}
\caption{Wall-normal mean profile of the (top row) streamwise rms velocity fluctuation, (middle row) wall-normal rms velocity fluctuation, and (bottom row) cross term of the Reynolds stress tensor $u'v'$. The three columns correspond to different sections: (left) $z=0$, (middle) $z=0.3h$, and (right) $z=0.6h$. The line and color style is the same as in \figrefA{fig:loglawE}.}
\label{fig:reyE}
\end{figure}
\figrefAC{fig:reyE} shows the profiles of the root mean square of the velocity fluctuations and of the Reynolds shear stress component as a function of the wall normal distance $\tilde{y}$. We observe that the streamwise velocity fluctuation $u_{rms}$ is only slightly altered by the increase of $Wi$, except its peak value which displaces further away from the wall. On the other hand, both the in plane fluctuation $v_{rms}$ and the shear Reynolds stress component $\overline{u'v'}$ are strongly affected by the increase of $Wi$. In particular, both $v_{rms}$ and $\overline{u'v'}$ are reduced across the whole domain, as already found for the turbulent channel flows with polymers \citep{ptasinski_boersma_nieuwstadt_hulsen_van-den-brule_hunt_2003a,dubief_white_terrapon_shaqfeh_moin_lele_2004a}.

\begin{figure}
\centering
\includegraphics[width = 0.25\textwidth]{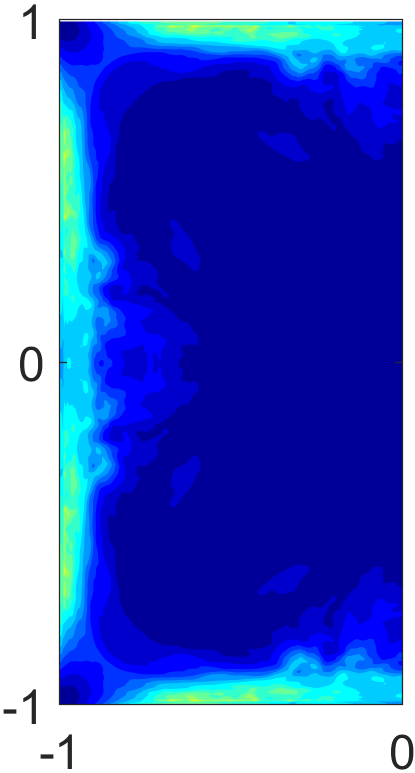}
\includegraphics[width = 0.25\textwidth]{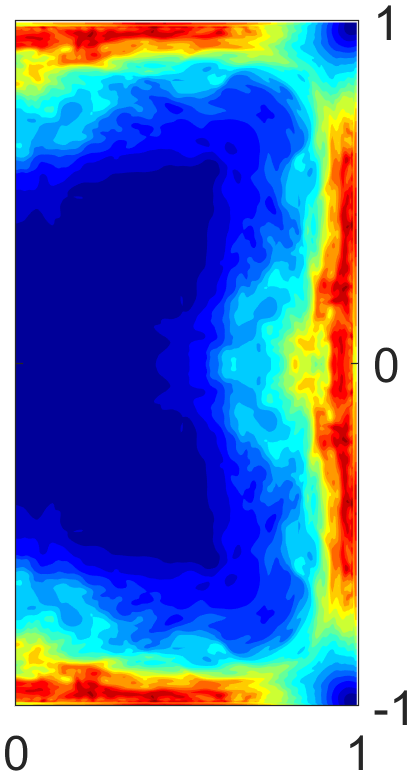}
\caption{Contour of the mean first normal difference $\mathcal{N}_1$, with the color scales ranging from $0$ (blue) to $0.8$ (red) for the polimeric cases with $Wi=1.5$ (left) and $Wi=3$ (right).}
\label{fig:normsE}
\end{figure}
Finally, we report in \figrefA{fig:normsE} the effect of the Weissenberg number on the mean first normal stress difference $\mathcal{N}_1$. This strongly increases with $Wi$, with its maximum value growing from approximately $0.3$ for $Wi=1.5$ (left panel) to $0.8$ in the high Weissenberg case (right panel). We also note that the spatial distribution remains similar for the different $Wi$ studied here, with the peak values of $\mathcal{N}_1$ close to the walls and decreasing to almost zero in the center of the duct and in the four corners. However, while for $Wi=1.5$ the region close to the centerline showed small values of the first normal stress difference, with a peak located at a distance of approximately $0.5h$ from the side walls, the case with $Wi=3$ exhibits a more uniform distribution around the wall bisectors, except in the corners where it remains almost null.

\section{Conclusion} \label{sec:conclusion}
We have performed numerical simulations of a turbulent duct flow with polymers, and compared the results with those of a Newtonian flow. The numerical simulations are direct numerical simulations, where all the space and time scales of the flow are resolved, and the presence of polymers is modeled with the FENE-P closure. The numerical simulations are performed at a fixed moderate Reynolds number $Re=2800$, resulting in a friction Reynolds number of $185$ for the Newtonian flow and $155$ for the viscoleastic counterpart with $Wi=1.5$, $\beta=0.9$ and $L^2=3600$; thus, the addition of polymers produces a drag reduction of approximately $30\%$ at the chosen Reynolds number. How the drag reduction is obtained and how it affects the secondary flow present in turbulent duct flows have been examined by different analysis.

We show that the mean streamwise velocity component is larger close to the corners in the viscoelastic flow than in the Newtonian case; the secondary flow is modified, with the locations of the maximum vorticity moving towards the center, away from the walls, and an increase of the circulation in each of the $8$ sectors. These effects are observed in the presence of polymers that strongly modifies the quasi streamwise vortices and the low- and high-speed streaks; in particular, their streamwise coherence is enhanced, they grow in size and depart from the walls, as already documented for turbulent plane channel flows with polymers. It is observed that, on average each edge hosts up to two high-speed streaks and one low-speed one, similarly to what found in duct flows at much lower bulk Reynolds number than the one considered here. However, we also found the flow to be deeply different from a Newtonian flow at the same friction Reynolds number.

The change in the near wall turbulent structures is accompanied by a modification of the turbulent fluctuations and stresses: the streamwise velocity fluctuations increase in the polymeric flow, whereas the in-plane components and the shear stresses are strongly reduced. In particular, it is found that the polymer force tends to reduce streamwise turbulent fluctuations in the bulk of the flow and in the four corners, while promoting those close to the walls, whereas the wall-normal fluctuations are always reduced except in the corners. This is further discussed in connection with the mean streamwise vorticity budget, related to the cross-flow velocity. The analysis reveals that although the polymeric contribution has a very small amplitude, the presence of polymers in the flow produces macroscopic changes in the flow itself; in particular, the contribution related to the in-plane derivatives of $\overline{v'w'}$ changes sign and becomes a production term, rather than a sink for the secondary motion.

The streamwise perturbation energy budget is also used to discuss the flow modification in the presence of polymers: again, we show that the viscoelastic flow is less uniform than its Newtonian counterpart, with the streamwise turbulent activity displaced farther away from the corners, towards the centerline. Also, the direct contribution of the polymer stress term in the energy budget is small in amplitude, but the polymer addition substantially modifies the flow itself, as for the streamwise vorticity budget.

We investigate the effect of the different parameters defining the viscoelastic behavior of the fluid, and show that the solution significantly changes with $Wi$. As  the Weissenberg number is increased, the  cross-stream vortical structures keep increasing in size, with their center further away from the wall. The slope of the inertial range strongly increases with $Wi$ and the mean velocity profile approaches the Maximum Drag Reduction asymptote. All the in-plane velocity fluctuations are strongly reduced by an increase of $Wi$, except for the streamwise component, which on the contrary remains almost unaltered.

\section*{Acknowledgment}
This work was supported by the European Research Council Grant no. ERC-2013-CoG-616186, TRITOS and by the Swedish Research Council Grant no. VR 2014-5001. The authors acknowledge computer time provided by SNIC (Swedish National Infrastructure for Computing).

\oneappendix%{Derivation of the vorticity budget} \label{sec:appendix}
\section{Derivation of the vorticity budget} \label{sec:appendix}
In this appendix we derive the equation for the mean streamwise vorticity component $\overline{\omega}_x$,
\begin{equation}\label{eq:a0}
\overline{\omega}_x = \frac{\partial \overline{w}}{\partial y} - \frac{\partial \overline{v}}{\partial z}.
\end{equation}
Using the Reynolds decomposition, the instantaneous flow variables are written as the sum of their mean value and fluctuation, \ie
\begin{equation}\label{eq:a1}
 u_i=\overline{U}_i+u'_i \;\;\;\;\;  p=\overline{p}+p' \;\;\;\;\; \tau_{ij}=\overline{\tau}_{ij}+\tau'_{ij},
\end{equation}
so as to obtain the following mean momentum equation
\begin{equation}\label{eq:a2}
\frac{\partial \overline{u}_i \overline{u}_j}{\partial x_j} +\frac{\partial \overline{u'_i u'_j}}{\partial x_j} = - \frac{\partial \overline{p}}{\partial  x_i} + \frac{\beta}{Re} \frac{\partial^2 \overline{u}_i}{\partial x_j \partial x_j} + \frac{1-\beta}{Re} \frac{\partial \overline{\tau}_{ij}}{\partial x_j},
\end{equation}
where we have considered that the mean flow is stationary. The streamwise component of the curl of \equref{eq:a2} then reads 
\begin{multline}\label{eq:a3}
\frac{\partial}{\partial y} \left( \frac{\partial \overline{w}~\overline{u}_j}{\partial x_j} +\frac{\partial \overline{w' u'_j}}{\partial x_j} \right) - \frac{\partial}{\partial z} \left( \frac{\partial \overline{v}~\overline{u}_j}{\partial x_j} +\frac{\partial \overline{v' u'_j}}{\partial x_j} \right) = \\
\frac{\beta}{Re} \frac{\partial^2}{\partial x_j \partial x_j} \left( \frac{\partial \overline{w}}{\partial y} - \frac{\partial \overline{v}}{\partial z} \right) + \frac{1-\beta}{Re} \left( \frac{\partial^2 \overline{\tau}_{3j}}{\partial x_j \partial y} - \frac{\partial^2 \overline{\tau}_{2j}}{\partial x_j \partial z} \right).
\end{multline}
Note that, the Schwarz's theorem has been used. By using the definition of streamwise vorticity (\equref{eq:a0}), the homogeneity in the streamwise $x$-direction ($\frac{\partial}{\partial x}=0$), and the incompressibility constraint for the mean flow ($\frac{\partial \overline{v}}{\partial y} + \frac{\partial \overline{w}}{\partial z} = 0$) one can write
\begin{multline}\label{eq:a5}
\overline{v} \frac{\partial \overline{\omega}_x}{\partial y} + \overline{w} \frac{\partial \overline{\omega}_x}{\partial z} + \frac{\partial}{\partial y} \left( \frac{\partial \overline{w' v'}}{\partial y} + \frac{\partial \overline{w' w'}}{\partial z} \right) - \frac{\partial}{\partial z} \left( \frac{\partial \overline{v' v'}}{\partial y} + \frac{\partial \overline{v' w'}}{\partial z} \right) = \\
\frac{\beta}{Re} \left( \frac{\partial^2}{\partial y^2} + \frac{\partial^2}{\partial z^2} \right) \overline{\omega}_x + \frac{1-\beta}{Re} \left( \frac{\partial^2 \overline{\tau}_{32}}{\partial y^2} + \frac{\partial^2 \overline{\tau}_{33}}{\partial z \partial y} - \frac{\partial^2 \overline{\tau}_{22}}{\partial y \partial z} - \frac{\partial^2 \overline{\tau}_{23}}{\partial z^2} \right).
\end{multline}
Finally, the equation can be rearranged to obtain \equref{eq:vorticityBalance}:
\begin{multline} \label{eq:a6}
\overline{v} \frac{\partial \overline{\omega}_x}{\partial y} +\overline{w} \frac{\partial \overline{\omega}_x}{\partial z} + \frac{\partial^2}{\partial y \partial z} \left( \overline{w' w'} -\overline{v' v'} \right) + \left( \frac{\partial ^2}{\partial y^2} - \frac{\partial ^2}{\partial z^2} \right) \overline{v'w'} = \\
\frac{\beta}{Re} \left( \frac{\partial ^2}{\partial y^2} + \frac{\partial ^2}{\partial z^2} \right) \overline{\omega}_x + \frac{1 - \beta}{Re} \left( \frac{\partial ^2 \overline{\tau}_{32}}{\partial y^2} + \frac{\partial ^2 \overline{\tau}_{33}}{\partial z \partial y} - \frac{\partial ^2 \overline{\tau}_{22}}{\partial y \partial z} - \frac{\partial ^2 \overline{\tau}_{23}}{\partial z^2} \right).
\end{multline}

%\section*{References}
\bibliographystyle{jfm}
\bibliography{./bibliography.bib}

\end{document}